\documentclass[aps, prb, english, notitlepage, twocolumn, superscriptaddress, floatfix]{revtex4-2} 
\usepackage{bm, color}
\usepackage{placeins}
\usepackage{nameref}
\usepackage{amssymb}
\usepackage{graphicx}
\usepackage{subfigure}
\usepackage{dcolumn}
\usepackage{bm}

\usepackage{mathtools, nccmath}

\allowdisplaybreaks
\usepackage[colorlinks = true,
            allcolors  = blue]{hyperref}
\usepackage{color,soul}

\def\H{\mathcal{H}}

\usepackage{CJK}
\usepackage{hyperref}

\preprint{APS/123-QED}
\begin{document}


\title{Transformer-based operator learning framework for self-energy in strongly correlated systems}

\author{Yuanran Zhu}
\email{yzhu4@lbl.gov}
\affiliation{Applied Mathematics and Computational Research Division, Lawrence Berkeley National Laboratory, Berkeley, California 94720, USA}

\author{Peter Rosenberg}%
\affiliation{Center for Computational Quantum Physics, Flatiron Institute, 162 5th Avenue, New York, NY 10010, USA}

\author{Zhen Huang}
\affiliation{Department of Mathematics, University of California, Berkeley, California 94720, USA}

\author{Hardeep Bassi}
\affiliation{Department of Applied Mathematics, University of California, Merced, California 95343, USA}

\author{Chao Yang}%
\affiliation{Applied Mathematics and Computational Research Division, Lawrence Berkeley National Laboratory, Berkeley, USA, 94720.}

\author{Shiwei Zhang}%
\affiliation{Center for Computational Quantum Physics, Flatiron Institute, 162 5th Avenue, New York, NY 10010, USA}

\begin{abstract}
We introduce $\Sigma$-Attention, a Transformer-based operator-learning framework for approximating the self-energy operator of strongly correlated electronic systems. By creating a batched dataset that combines results from three complementary approaches: many-body perturbation theory, strong-coupling expansion, and exact diagonalization, each effective in specific parameter regimes, $\Sigma$-Attention is applied to learn an accurate approximation for the self-energy operator that is valid across a wide range of parameter regimes. This hybrid strategy leverages the strengths of existing methods while relying on the transformer's ability to generalize beyond individual limitations. More importantly, the scalability of the Transformer architecture allows the learned self-energy to be extended to systems with larger sizes, 
leading to much improved computational scaling.
Using the 1D Hubbard model, we demonstrate that $\Sigma$-Attention can accurately predict the Matsubara Green's function of large systems with a wide range of coupling strength. Our framework offers a promising and scalable pathway for studying strongly correlated systems
with many possible generalizations.
\end{abstract}

\maketitle

\section{Introduction}

Strongly correlated materials have long fascinated the condensed matter community due to their rich quantum phenomena, including high-temperature superconductivity, magnetism, and Mott transitions. These emergent properties arise from intricate electron--electron interactions that often render conventional single-particle approaches such as density functional theory (DFT) inadequate. As a result, accurately modeling strongly correlated systems remains one of the most challenging problems in theoretical 
and computational physics and chemistry.

Over the years, a variety of methods have been developed to study strongly correlated systems. These include dynamical mean-field theory (DMFT) \cite{georges1996dynamical,kotliar2006electronic,kotliar2001cellular}, strong-coupling expansion (SCE) and cluster perturbation theory (CPT) \cite{pairault1998strong,pairault2000strong,metzner1991linked,stanescu2004strong,gros1993cluster,senechal2000spectral,senechal2002cluster}, quantum cluster methods \cite{maier2005quantum,hettler2000dynamical,macridin2006phase}, various quantum Monte Carlo (QMC) techniques \cite{zhang1997constrained,zhang2003quantum,gull2011continuous,van2010diagrammatic,rossi2017determinant,antipov2017currents,blankenbecler1981monte}, 
and tensor-based approaches such as the density matrix renormalization group (DMRG) and its extensions \cite{schollwock2005density,liu2025accurate,chan2011density,chan2016matrix}. All of these approaches provided valuable insights into different systems and have shown to be effective in different parameter regimes. However, each method has inherent limitations. For instance, SCE is typically restricted to very strong interactions, DMRG does not scale as well in extended systems,
and QMC loses exactness in the presence of the sign problem.
These challenges motivate the search for a unified framework capable of capturing the complex behavior across all regimes.

In recent years, the rapid development of machine learning (ML)—particularly deep learning—has opened new avenues for addressing complex many-body problems. Notable advances include neural-network-based representations of many-body/Kohn–Sham states or DFT functional \cite{torlai2018neural,jia2019quantum,pfau2020ab,hou2024unsupervised,nelson2019machine}, efficient calculation of two-electron integrals \cite{liang2024effective,liang2025exploring}, and Green's function-based approaches for studying many-body interactions \cite{bassi2024learning,zhu2025predicting,kakizawa2024physics,agapov2024predicting,venturella2024unified}. In these studies, various ML models—such as convolutional neural networks (CNNs), variational autoencoders, and recurrent neural networks (RNNs)—have been employed to model quantum states and calculate observables. Among these, we are particularly interested in ML models that use the Transformer architectures. Originally developed for natural language processing \cite{vaswani2017attention,ainslie2020etc,lu2024deepseek}, Transformer has demonstrated remarkable success in learning operators \cite{kissas2022learning,calvello2024continuum, hao2023gnot,boulle2023mathematical} and capturing long-range dependencies. Their inherent scalability 
and ability to handle variable input dimensions make them an attractive tool for studying strongly correlated systems, especially for addressing 
finite-size effects and improving computational scaling towards the thermodynamic limit with
simulations.

Driven by these motivations, in this work, we introduce \(\Sigma\)-Attention, a transformer-based operator-learning framework designed to approximate the self-energy operator of strongly correlated electronic systems. Leveraging an Encoder-Only transformer as our ansatz, our approach learns a mapping that predicts the self-energy \(\Sigma(k,\mathrm i\omega_n)\) from the non-interacting Green's function \(G_0(k,\mathrm i\omega_n)\) and the two-body interaction \(v\). A distinctive aspect of our method is the use of a batched dataset that integrates complementary data from many-body perturbation theory (MBPT), strong-coupling expansion (SCE), and exact 
results from smaller sizes, e.g., by diagonalization (ED), each of which is effective in specific parameter regimes. This enables the transformer to learn an accurate approximation to the self-energy operator that remains valid across a wide range of parameter regimes. Consequently, our hybrid strategy combines the strengths of these established methods and exploits the transformer's generalization capability to extend predictions to systems with varying sizes and interaction strengths. We test the utility of this framework using a schematic 1D Hubbard model, where accurate numerical results in large system sizes are obtained by auxiliary-field quantum Monte Carlo \cite{zhang1997constrained,zhang2003quantum} calculations. We demonstrate
that \(\Sigma\)-Attention yields high-fidelity predictions of the Green's function for systems with different sizes and interaction strengths, 
accurately capturing
the metal-to-insulator transition over a wide range of \(U\) values.

The remainder of this paper is organized as follows. In Section~\ref{sec:pre}, we present the preliminary
and
background necessary 
for this study. Section~\ref{sec:Method} details the \(\Sigma\)-Attention framework, including its architecture and training strategies. In Section~\ref{sec:app}, we apply our method to the 1D Hubbard model and compare the predictions with benchmark results. Finally, Section~\ref{sec:Conclusion} summarizes the main findings and outlines future research directions. Section~\ref{sec:details} provides all technical details for the numerical simulations.

\section{Preliminary}
\label{sec:pre}
Consider a generic fermionic quantum many-body system in a lattice:
\begin{equation}\label{MB_ham}
    \mathcal{H}  = \sum_{ij,\sigma}t_{ij}c^\dagger_{i\sigma}c_{j\sigma} + \frac{1}{2}\sum_{ijkl}\sum_{\sigma\sigma'} v_{ijkl}^{\sigma\sigma'} c^\dagger_{i\sigma}c_{j\sigma'}^\dagger c_{k\sigma'} c_{l\sigma},
\end{equation}
where $t_{ij}$ represents the single-particle Hamiltonian (including kinetic energy and any external potential) and $v_{ijkl}^{\sigma\sigma'} $
are the two-electron integrals encoding the interaction between electrons. The equilibrium properties of this interacting system are captured by the Matsubara Green's functions. Focusing on single-particle excitations, the single-particle Matsubara Green's function is defined as
\[
G_{ij,\sigma}(\tau) = -\langle \mathcal T c_{i\sigma}(\tau) c_{j\sigma}^\dagger(0) \rangle, 
\]
where \(\mathcal T\) is the time-ordering operator in imaginary time and $\langle\cdot\rangle:=\text{Tr}[e^{-\beta\H}]/Z$ denotes the ensemble average. By performing a Fourier transformation with respect to the imaginary time $\tau$, one obtains the Green's function in frequency space $G_{ij,\sigma}(  \mathrm i\omega_n)$ which is defined on the discrete Matsubara frequencies $  \mathrm i\omega_n=(2n+1)\pi/\beta$. In this study, we focus on the system with equal spin-up and spin-down electrons
and hereafter ignore the spin indices in the Green's function. 
The following discussion is valid for both the lattice Green's function $G_{ij}(\tau), G_{ij}(\mathrm i\omega_n)$ and the $k$-space Green's function $G(k,\tau), G(k,\mathrm i\omega_n)$ commonly used for periodic systems. Hence we will use simple notations $G, G_0,\Sigma$ instead and write down their explicit expressions whenever needed. 

Dyson’s equation connects the non-interacting Green’s function \(G_0\), to the full (interacting) Green’s function \(G\), through the self-energy \(\Sigma\), which encapsulates all the effects of interactions. In frequency space, Dyson’s equation is typically written as
\begin{align}
\label{dyson_eqn}
G(\mathrm i\omega_n) = G_0(\mathrm i\omega_n) + G_0(\mathrm i\omega_n)\, \Sigma(\mathrm i\omega_n)\, G(\mathrm i\omega_n),
\end{align}
or, equivalently, in its inverse form,
\begin{align}\label{dyson_eqn_inv}
G^{-1}(\mathrm i\omega_n) = G_0^{-1}(\mathrm i\omega_n) - \Sigma(\mathrm i\omega_n).
\end{align}
By determining the self-energy \(\Sigma(\mathrm i\omega_n)\), one can solve Dyson's equation to obtain the interacting Green's function, from which various physical observables can be derived. For example, the density of states (DOS) $\rho(\omega)$, which is key to identifying phenomena such as the metal-to-insulator transition, is computed by first performing an analytic continuation \(\mathrm i\omega_n\to\omega+i0^+\) \cite{fei2021analytical, huang2023robust} to obtain the retarded Green's function \(G^R(\omega)\), and then using the relation
\begin{align}\label{DOS}
\rho(\omega) = -\frac{1}{\pi}\operatorname{Im}G^R(\omega).
\end{align}
For weakly correlated systems in which \(|v_{ijkl}|\ll |t_{ij}|\), many-body perturbation theory (MBPT) is normally employed to approximate the self-energy. In the bare expansion approach, the self-energy is viewed as an operator that maps the non-interacting Green's function \(G_0\) and the two-body interaction \(v\) (with \(v\) used as shorthand for \(v_{ijkl}\)) to the self-energy \(\Sigma\). This mapping is expressed via the expansion series
\[
\Sigma(\mathrm i\omega_n) = \Sigma[v,G_0](\mathrm i\omega_n) = \underbrace{\Sigma^{(1)}(\mathrm i\omega_n)}_{O(|v|)} + \underbrace{\Sigma^{(2)}(\mathrm i\omega_n)}_{O(|v|^2)} + \cdots,
\]
where the \(n\)th-order contribution \(\Sigma^{(n)}\) is represented by the corresponding Feynman diagrams. Resummation techniques can also be applied to obtain the renormalized self-energy operator \(\Sigma = \Sigma[v, G](\mathrm i\omega_n)\). 
Consequently, Dyson's equation~\eqref{dyson_eqn} must be solved self-consistently until convergence is reached.

Perturbative calculations typically fail for systems with intermediate or strong correlations. In such cases, bare approximations often yield non-physical results, while renormalized self-energy approaches—such as the second-order Born (2ndB) or GW methods—may simply fail to converge. Consequently, phenomena like the Mott transition cannot be captured by simple perturbation theory. To address these shortcomings, more advanced self-energy approximations have been developed, notably those based on Dynamical Mean-Field Theory (DMFT) \cite{georges1996dynamical} and its extensions, including Cellular DMFT (CDMFT) \cite{kotliar2001cellular} and the Dynamical Cluster Approximation (DCA)\cite{hettler2000dynamical,macridin2006phase}, which effectively incorporate short-range spatial fluctuations.

In general, DMFT-based approaches introduce a non-perturbative ansatz for the self-energy operator by matching the self-energy of the embedded impurity or dynamical clusters to the global lattice self-energy. In this work, we show that a transformer-based neural network can provide a novel ansatz, with the matching process ensuring that the ansatz is generally valid across different parameter regimes of the model.

\section{Method}
\label{sec:Method}
\subsection{Problem setup}
While the self-energy operator is traditionally derived using MBPT, the functional-integral approach \cite{potthoff2004non} indicates that, even in the non-perturbative regime, the self-energy can be expressed self-consistently as an functional of the Green's function:
\[
\Sigma = \Sigma[G] = \frac{\delta \Phi}{\delta G},
\]
where \(\Phi\) is the Luttinger-Ward functional. 
This suggests the existence of a \emph{universal} self-energy operator that maps the interactive Green's function \(G\) and the two-body interaction \(v\) onto \(\Sigma\). We further \emph{assume} that this universality is also valid for the bare Green's function $G_0$ 
\footnote{
    Technically speaking, this amounts to assume the existence of a functional $\tilde \Phi[G_0]$ such that $\Sigma =\delta \tilde \Phi/\delta G_0$. We emphasize that whether this is generally valid is an open question. In this work, we only use this ansatz to motivate the NN-representation of a universal self-energy operator valid both at weak and strong couplings.}
. The aim of this work is to find NN approximations to the self-energy operator:

\begin{align*}
G_0(\mathrm i\omega_n), v &\xrightarrow{\text{NN}} \Sigma(\mathrm i\omega_n):\quad \text{(Bare)}\\
G(\mathrm i\omega_n), v &\xrightarrow{\text{NN}} \Sigma(\mathrm i\omega_n):\quad \text{(Renormalized)}
\end{align*}   
where the $\tau$-space formulation can be similarly defined. In this work, we only test the bare ansatz since the subsequent computation of \(G\) from the NN-predicted self-energy is straightforward, requiring no self-consistent iterations. 
The postulated universality would require the approximated self-energy operator to be valid across systems of various sizes, for a particular form of two-body interaction (e.g., Coulomb) \(v\), 
and for any \(G_0(\mathrm i\omega_n)\). However, achieving this complete universality using NN is nearly impossible and, in many cases, unnecessary, as the physical properties of electronic systems can vary significantly depending on factors such as geometry, dispersion relations, and  interaction strength. Therefore, we will focus our discussions on the 1D Hubbard model. Our goal with machine learning is to design an appropriate neural network and select a sufficiently large and representative dataset, so that, after proper training, the NN-approximated self-energy operator can generalize as broadly as possible to Hubbard models with different interaction strengths 
$U$ and varying system sizes. In particular, achieving generalizability across system sizes is highly desirable, as it addresses the challenging finite-size effect problems. With that being said, the approach we developed should be generally applicable to a generic electronic system \eqref{MB_ham}, given the right choice of training dataset.

In the following subsections, we discuss separately the design of the neural network and the selection of the training dataset, which are equally important in our construction and reflect the motivations highlighted above.
\subsection{Encoder-Only Transformer for operator learning}
\begin{figure}
\centering
\includegraphics[width=1.0\linewidth]{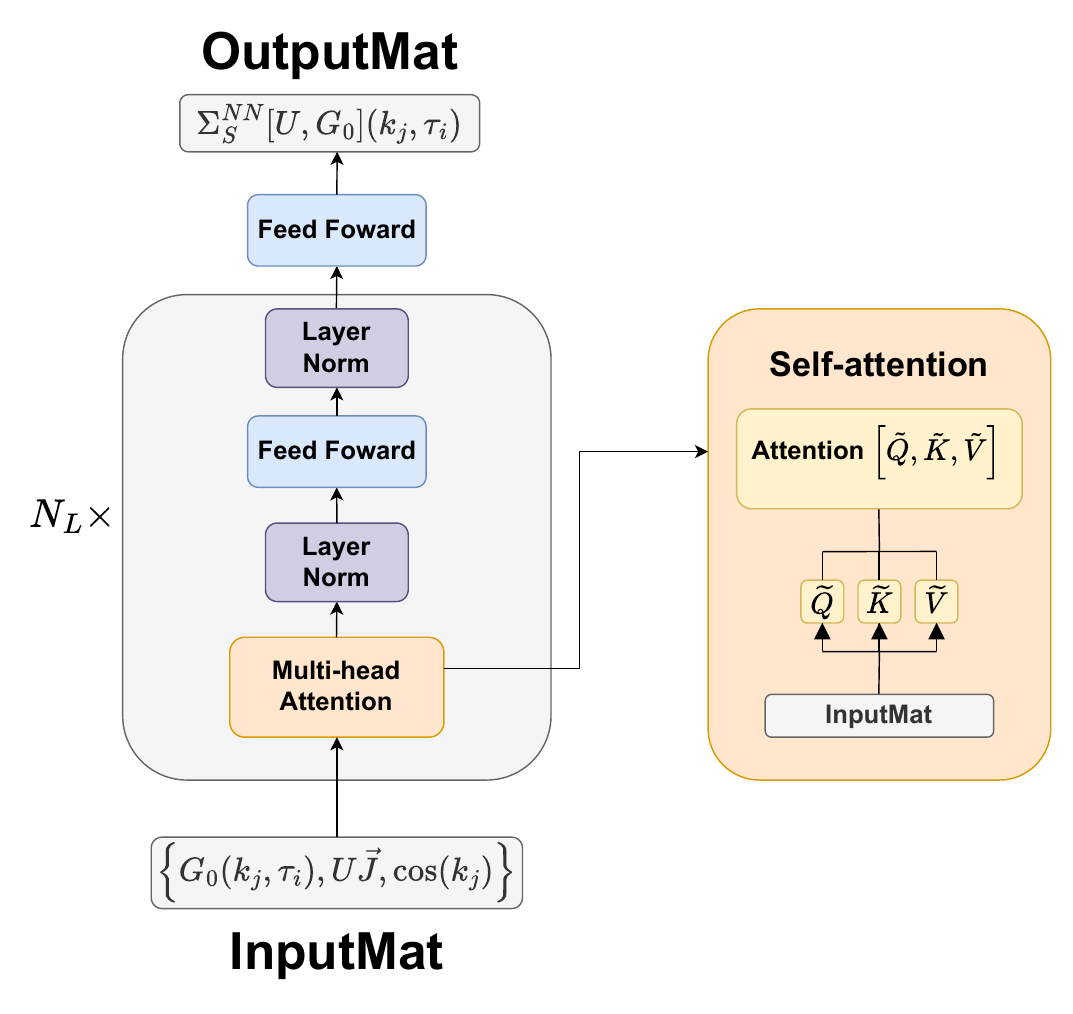}
\caption{Encoder-only transformer architecture employed as the neural network module for  \(\Sigma\)-Attention. The model requires the bare Green's function $G_0$, interaction strength $U$, and positional encoding $\cos(k)$ as input, which are concatenated into a single matrix $\text{InputMat}$ (see details in Section \ref{subsec:Learning and Training details}); Shown on the right is a schematic of a single-head self-attention module. For full details on the multi-head extension, see Section \ref{sec:details}. The \text{InputMat} is multiplied by different learnable weight matrices $W^Q, W^K, W^V$ to form $\tilde{Q}, \tilde{K}, \text{ and } \tilde{V}$, respectively. This module generates its output via (\ref{eqn:self-att}).  A fully connected (feed forward) layer is then used to project the output to the target dimension of the scaled self-energy matrix prediction $\Sigma_S^{NN}$. The loss between the predicted self-energy and ground truth self-energy is then measured to update the total network parameters $\theta$. Here, $\theta$ encompasses all parameters that parameterize the weight matrices in the self-attention block(s) and feed forward layer(s). Post-training, with the predicted $\Sigma_S^{NN}$, the Green's function \(G(k,\tau)\) is solved for via \eqref{dyson_eqn_inv}.}
\label{fig:sigma_att}
\end{figure}

To achieve system size generalizability, the neural network must be trainable on data $\{G_0,\Sigma\}$ from small systems and then be applicable to larger systems. Conventional neural network architectures, such as multilayer perceptrons (MLPs), are not well suited for this task because they require weight matrices and bias vectors whose dimensions depend on the input size. As the system size changes, the dimensionality of the modeling weight matrices and bias vectors must be changed accordingly.
This problem is well-addressed using Transformer models due to its dimensional-agnostic architecture. To see this, it is suffice to examine its core computational unit, namely the self-attention mechanism (see FIG. \ref{fig:sigma_att}), which has the following computational formula:
\begin{align}\label{eqn:self-att}
\text{Attention}[\tilde{ Q},\tilde{K},\tilde{V}]
=\sigma\!\left(\frac{\tilde{ Q}\tilde{K}^T}{\sqrt{n_f}}\right)\tilde{ V},
\end{align}
where the input is an $n_p \times n_f$ matrix (InputMat). This matrix is projected through learnable weight matrices $W^Q, W^K, W^V$ (dimension of which are $n_f \times n_d$, $n_f\times n_d$, and $n_f\times n_f$) to produce $\tilde{Q}, \tilde{K}, \tilde{V}$. For example, $\tilde{Q}= \text{InputMat} \cdot W^Q$. The crucial observation is that the dimensions of these weight matrices depend only on the {\em fixed} feature dimension $n_f$ and hidden dimension $n_d$, but \textit{not} on the physical dimension $n_p$. Consequently, the same set of weight matrices can process inputs of any physical dimension $n_p$, making the architecture independent of system size.

For our self-energy prediction task, let us give a simple example of using the above one-layer self-attention as the modeling ansatz for the operator. For this setting, the input matrix InputMat has dimensions $N_k \times (N_\tau + 2)$, composed of three components: (i) the bare Green's function values $G_0(k_i, \tau_j)$, (ii) the interaction strength $U$ (broadcasted as a uniform column), and (iii) positional encodings $\cos(k_i)$. Here, $k_i$ ($i=1,2,\ldots,N_k$) and $\tau_j$ ($j=1,2,\ldots,N_\tau$) represent uniformly sampled momentum and imaginary-time grid points on an equally spaced grid with $N_\tau = 101$ points (see Section \ref{subsec:Learning and Training details} for details of the positional encodings).
In this example, the physical dimension $n_p = N_k$ varies with system size, while the feature dimension $n_f = N_\tau + 2$ and the hidden dimension $n_d$ is a hyperparameter of the NN that remain fixed across all systems. Because the self-attention weight matrices depend only on $n_f$ and $n_d$, a single trained model can accept Green's functions from systems of any size. The self-attention output only needs to pass through a linear projection (multiplying by a matrix of dimension $n_f \times (n_f-2)$) to produce the final predictions of shape $N_k \times N_\tau$, which correspond directly to the self-energy $\Sigma(k_i, \tau_j)$ for all $k$ and $\tau$ grids.  In specific implementation, we actually use the Multi-Head Attention as the modeling ansatz for the self-energy. Nevertheless, the above analysis for the dimension-agnostic feature of the Transformer architecture still applies. The full architectural details, including the definition of Multi-Head Attention, training procedures, and data handling, are provided in Section \ref{subsec:Learning and Training details}.

From an operator-learning perspective, the self-energy operator is approximated using the Transformer ansatz. The network is trained by minimizing the difference $\|\Sigma^{NN}[U, G_0] - \Sigma[U, G_0]\|$ across a variety of input pairs $\{G_0, U\}$. Since the exact self-energy for arbitrary $U, G_0$ is unknown, the training uses data where reliable approximations (from MBPT, SCE, or ED) are available. The trained network then recovers the self-energy operator in regimes where exact solutions are difficult to obtain.
%
%
\subsection{Complementary datasets for training NN}
\begin{figure}
\centering
\includegraphics[width=7cm]{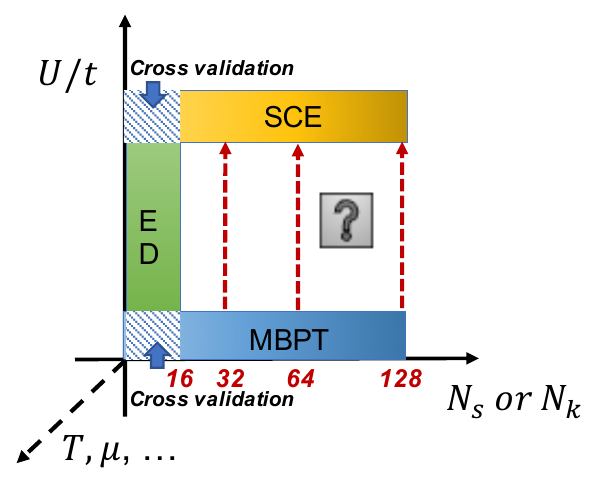}
\caption{Complementary datasets used for training \(\Sigma\)-Attention. In the shaded overlapping regime, two different approaches can be used to generate the same training data, allowing cross-validation and establishing the applicability range of perturbative methods such as MBPT and SCE. The area marked by a question mark (corresponding to large \(N_s\) and intermediate interaction strength) represents a regime that is typically challenging for current numerical methods, which we aim to explore using the generalizability of \(\Sigma\)-Attention.
This paper only considers the half-filling case and fixed temperature. Nevertheless, system parameters such as temperature \(T\) and filling \(\mu\) could be incorporated into the dataset to study \(T\)- or \(\mu\)-induced 
phase transition or crossover. This awaits further investigations.}

\label{fig:data_structure}
\end{figure}
The Transformer architecture offers promising system-size scalability for representing the self-energy operator. However, achieving generalizability with respect to the interaction strength \(U\)---that is, ensuring that the self-energy learned at certain values of \(U\) can be extended to other values---is nontrivial. Relying solely on MBPT data is clearly insufficient for this purpose. To capture phenomena such as the metal-to-insulator transition, it is essential to provide the neural network with data spanning both weakly and strongly correlated regimes. This motivates the use of the following complementary datasets for training the neural network, each of 
which is generated by a different approach. The applicability ranges of these methods are summarized in FIG \ref{fig:data_structure}. All the calculation details are provided in Section \ref{subsec:data_source}.
\paragraph{MBPT} For weakly correlated systems where \(U \ll t\), many-body perturbation theory (MBPT) using self-energy approximations such as the second-order Born (2ndB) and GW methods \cite{stefanucci2013nonequilibrium} provides quantitatively accurate estimates of both the Green's function and the self-energy. Owing to the perturbative nature of MBPT, the computational cost of generating the training data \(\{G_0, \Sigma^{\text{2ndB/GW}}\}\) remains relatively low even for large systems. Consequently, as illustrated in FIG. \ref{fig:data_structure}, the MBPT data occupies the lower rectangular regime of the overall data space. 
\paragraph{SCE} In the regime of extremely strong correlations, where \(U \gg t\), a perturbative approach known as the strong-coupling expansion (SCE) \cite{pairault1998strong,dupuis2000strong,pairault2000strong,metzner1991linked,stanescu2004strong} can be employed to obtain the Green's function by perturbing around the atomic limit. Unlike the weak-coupling MBPT, the SCE regards the Hubbard-\(U\) term as the non-perturbative core and expands in powers of $t_{ij}$. Consequently, this formulation does not directly approximate the conventional self-energy defined in \eqref{dyson_eqn}, but rather its strong-correlation analogue \(\Gamma(\mathrm i\omega_n)\). In practice, we use the SCE to compute the interacting Green's function \(G\) and then solve \eqref{dyson_eqn_inv} to extract the corresponding self-energy \(\Sigma^{\text{SCE}}\) \cite{pairault1998strong,dupuis2000strong,pairault2000strong}. SCE is also a perturbative theory and hence occupies the upper rectangular regime in the overall data space as illustrated in FIG \ref{fig:data_structure}.
\paragraph{ED} In the regime of intermediate interaction strength, perturbation theory often struggles, necessitating the use of alternative methods. Common approaches include quantum Monte Carlo (QMC)\cite{qin2016benchmark,zhang1997constrained,zhang2003quantum,gull2011continuous}, density matrix renormalization group (DMRG)\cite{white1998strongly,schollwock2005density}, and exact diagonalization (ED)\cite{fulde1995electron,iskakov2018exact}. These methods can yield quantitatively accurate results within their respective domains, yet each has its limitations. Compared to perturbative techniques, these methods are computationally more demanding and are thus often limited to relatively small systems. As depicted in FIG. \ref{fig:data_structure}, they occupy the left rectangular regime of the data space. 
In this work, 
we utilize an ED solver \cite{iskakov2018exact} to generate this portion of the dataset. Exact results can be produced for larger system sizes in this particular case, for example using AFQMC. However, since our goal in this work is to develop and test the operator learning approach, we leave them for use in validation only.

All these complementary datasets are combined into a single training dataset to devise $\Sigma$-Attention to learn a global self-energy operator that is valid for any interaction strength \(U\). As we can see, the assembly of the training data in our design is inherently \emph{modular}; that is, any component of the combined dataset can be replaced if an alternative approach yields quantitatively accurate results within a given parameter range. For example, at half-filling, AFQMC can substitute for other methods since the fermionic sign problem is absent. Likewise, cluster perturbation theory \cite{senechal2002cluster,senechal2000spectral} is a viable alternative to the strong-coupling expansion. The guiding principle is that any alternative or new method must provide quantitatively accurate results---otherwise, incorporating inaccurate data would contaminate the data pool and lead to suboptimal training outcomes (see Appendix \ref{app:importance_dataset} for further discussion). Additionally, the generation of these datasets should be computationally efficient, remaining within the comfort zone of the corresponding method, so that the benefits of training a neural network for generalization are fully realized.

On the other hand, \(N_k\) and \(U\) are not the only system parameters that determine the applicability range of different approaches. In principle, our framework can be generalized to predict and characterize phase transitions associated with any order parameters
\footnote{Presumably, the phase transition or crossover under the investigation of $\Sigma$-Attention should be at least second order. Otherwise, the network will likely generate a smooth approximation to the change of order parameter cross the critical point.}
, such as temperature \(T\) and filling \(\mu\). The corresponding bare ansatz would then be
\[
G_0(\mathrm i\omega_n),U, T, \mu,\, \cdots \xrightarrow{\text{NN}} \Sigma(\mathrm i\omega_n) \quad \text{(Bare)}.
\]
Accordingly, one may consider incorporating additional datasets, such as those generated by high-temperature expansion \cite{bartkowiak1995high,perepelitsky2016transport}. This, however, will not be addressed in this paper and is left for future work.

\section{Application to Hubbard model}
\label{sec:app}
\begin{figure*}
\centering
\includegraphics[width=18cm]{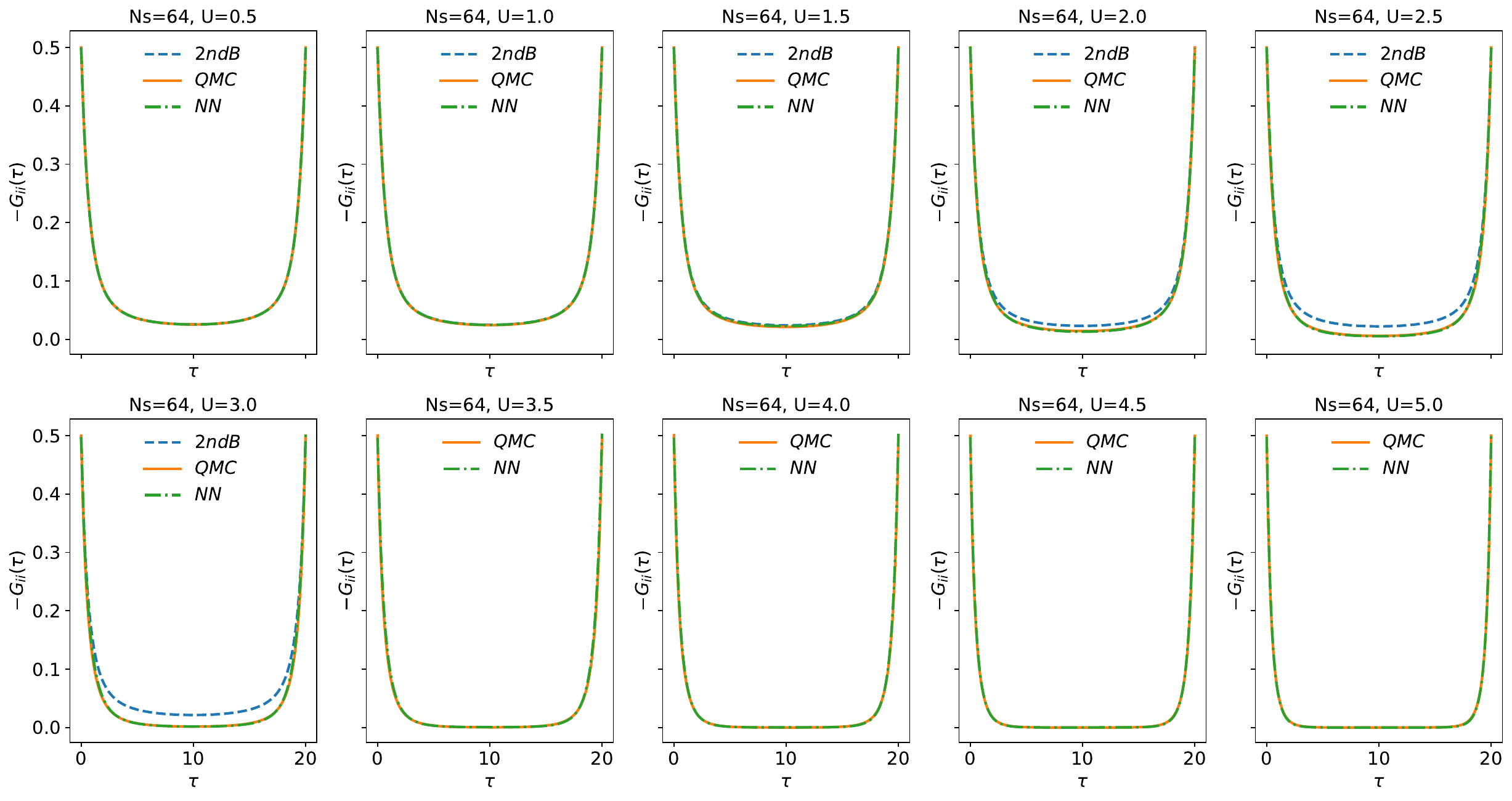}
\caption{Comparison of the onsite Green's function $G_{ii}(\tau)$ for Hubbard model with $N_s=64$ computed using AFQMC, $\Sigma$-Attention, and second-order Born (2ndB) calculations. Note that due to the PBC, $G_{ii}(\tau)$ is the same for any site $i$. The self-consistent 2ndB calculation does not converge for $U>3.0$, therefore we only displayed the results up to $U=3.0$.}
\label{fig:Gii}
\end{figure*}
\begin{figure*}
\centering
\includegraphics[width=18cm]{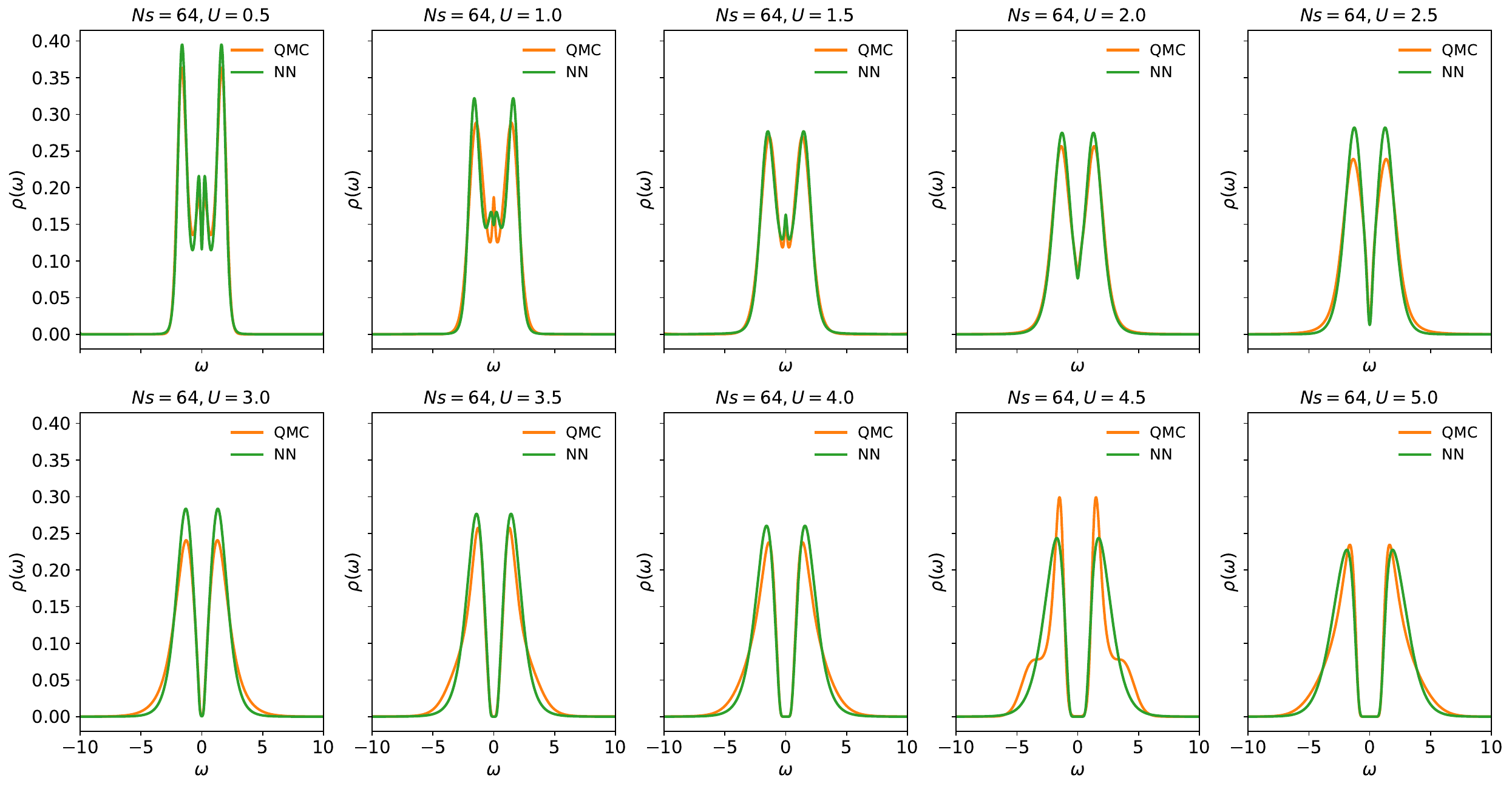}
\caption{Comparison of denisty of state (DOS) $\rho(\omega)$ for Hubbard model with $N_s=64$ computed using the Green's function $G_{ii}(\tau)$ obtained by AFQMC and $\Sigma$-Attention. We see that the spectral gap opening is well-captured by $\Sigma$-Attention.}
\label{fig:DOS}
\end{figure*}
We apply $\Sigma$-Attention to 1D Hubbard model at half-filling to demonstrate the effectiveness of this operator learning approach. In particular, we focus on whether the model trained on the MBPT-SCE-ED combined datasets can be generalized to predict Green's function of large systems with intermediate interaction strength where the metal-to-insulator transition happens. The modeling Hamiltonian is given by
\begin{align}\label{hubbard}
    \H =-t\sum_{\langle ij\rangle\sigma}c_{i\sigma}^{\dagger}c_{j\sigma}+\sum_{i}Un_{i\uparrow}n_{i\downarrow}-\mu\sum_{i\sigma}n_{i\sigma}
\end{align}
where $\langle ij\rangle$ denotes the nearest-neighbor hopping, the chemical potential is set to be $\mu=U/2$, corresponding to the half-filling case. The periodic boundary condition is imposed for $N_s =4,8,12,\cdots$, while the anti-periodic boundary condition is imposed for $N_s =6,10,14,\cdots$ 
\footnote{The anti-PBC is chosen to avoid the finite-size effect for system sizes $N_s =6,10,14,\cdots$ since otherwise the metal-to-insulator transition cannot be seen due to this sampling of $k$-points in the first Brillouin zone.}.
The inverse temperature is fixed to be $\beta=20$. We should note that, while in this paper, we only test $\Sigma$-Attention in Hubbard model, the framework is generally applicable to systems with general two-body interaction $v_{ijkl}$.
\begin{table}[t]
    \centering
    \begin{tabular}{ccc}    
        \textbf{Method} & \(\boldsymbol{N_s}\) & \(\boldsymbol{U/t}\)\\[0.5ex]
        \hline\hline
        ED   & 4, 6, 8, \(\cdots\), 16  \hspace{0.2cm}     & 0.5, 1.0, \(\cdots\), 7.0 \\
        MBPT-I& 32, 34, \(\cdots\), 128 \hspace{0.2cm}         & 0.2, 0.3, \(\cdots\), 1.5 \\
        MBPT-II & 20, 24, 28 \hspace{0.2cm}         & 0.2, 0.3, 0.4, 0.5 \\
        SCE  & 20, 34, \(\cdots\), 128\hspace{0.2cm}          & 6.0, 6.5, 7.0 \\
    \end{tabular}
    \caption{System parameters for each method used for generating the training data for $\Sigma$-Attention. Note that in the overlapping regime, MBPT and SCE results are benchmarked with ED to determine their validity ranges. Details can be found in Section \ref{subsec:Learning and Training details}. When $20\leq N_s<32$, the finite-size effect will lead to inaccuracy of the 2ndB calculation for relatively large $U$. Hence we split the MBPT datasets into two parts as shown above.}
    \label{table:dataset}
\end{table}
\subsection{Numerical results}
Table \ref{table:dataset} summarizes the training dataset used in our study. After proper training of \(\Sigma\)-Attention (see Section \ref{subsec:Learning and Training details} for details), we test whether the network can accurately predict the Green's function dynamics for large systems with \(N_s = 32, 64, 128\) in the intermediate coupling regime. The training dataset and target prediction domains are illustrated in Figure \ref{fig:data_structure}.
Since the network takes \(G_0\) and \(U\) as input and outputs the self-energy, the full Green's function \(G(k,\mathrm i\omega_n)\) is computed by solving
\begin{align}\label{dyson_eqn_inv_NN}
G^{-1}(k,\mathrm i\omega_n) = G_0^{-1}(k,\mathrm i\omega_n) - \Sigma^{NN}[G_0,U](k,\mathrm i\omega_n).
\end{align}
Throughout training and prediction, \(G_0(k,\mathrm i\omega_n)\) is chosen to be the Hartree–Fock (HF) Green's function. Since \(\Sigma\)-Attention produces output almost instantly, computing \(G(k,i\omega_n)\) is extremely efficient and requires virtually no additional computational cost other than matrix inversion in \eqref{dyson_eqn_inv_NN}. After completing the above procedures, we also apply the adaptive pole-fitting techniques developed in~\cite{huang2023robust} to \(G(k, i\omega_n)\) to obtain its \emph{causal} approximation. This final step filters out numerical noise and ensures that the resulting Green’s function satisfies the required causality conditions.

\begin{figure}
\centering
\includegraphics[width=7cm]{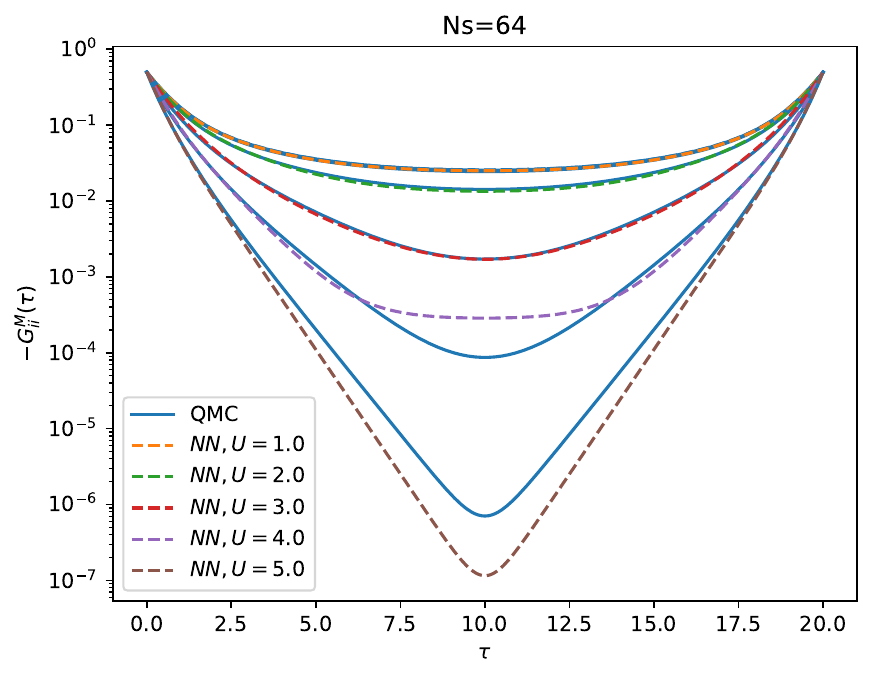}
\caption{Prediction of the onsite Green's function $G_{ii}(\tau)$ displayed on the log scale.}
\label{fig:Gii_log}
\end{figure}

FIG \ref{fig:Gii} presents the final calculation result of \(G_{ii}(\tau)\) for \(N_s = 64\) and \(U = 0.5, 1.0, \ldots, 5.0\). For benchmarking, we compare our results with AFQMC and MBPT outcomes where the later is obtained using the 2ndB self-energy approximation. At half-filling, AFQMC is sign-problem free, making it an ideal benchmark. As shown in the figure, \(\Sigma\)-Attention yields accurate predictions for the Green's function across all \(U\) values, whereas the 2ndB approximation deteriorates as \(U\) increases, as expected. The comparison between AFQMC results and the NN prediction is shown more clearly in FIG. \ref{fig:Gii_log}. The overall prediction error,
\[
\|G_{ii}^{NN}(\tau) - G_{ii}^{QMC}(\tau)\|_\infty,
\]
for all the $U$ values is on the order of \(10^{-4}\) to \(10^{-2}\); similar error bound holds for the off-diagonal elements \(G_{ij}(\tau)\). The additional numerical results for Hubbard model with $N_s= 32, 128$ are  provided in Appendix \ref{app:add_numerial} and the conclusion is the same. 
To clearly see the metal-to-insulator transition predicted by $\Sigma$-Attention, we also compare the density of state (DOS) \eqref{DOS} obtained by the analytic continuation (AC) of the Matsubara Green's function using the Maximum entropy method \cite{kaufmann2023ana_cont}. 
The AC procedure is known to be ill-posed and the approximation error is inevitably amplified in $\rho(\omega)$. However, due to the high accuracy of the AFQMC results, the corresponding AC error is better controlled which makes it a reliable benchmark when comparing the main physical features of the spectral function. From FIG \ref{fig:DOS}, we clearly see that the NN-predicted DOS well captures the opening of the spectral gap and have a overall shape agrees with the AFQMC results.

Based on the obtained numerical results, $\Sigma$-Attention demonstrates great scalability and delivers accurate predictions of Green's function dynamics for systems as large as $N_s=128$ under intermediate interaction strengths. Comparatively, AFQMC simulations scale roughly as $\mathcal{O}(N_s^3)$ and $N_s = 128$ systems treated here therefore require modest computational resources, in the range of $\sim 300$ CPU core-hours.  In the current framework, the causality of the Matsubara Green’s function is enforced through a post-processing step that applies numerical causal projection. A more principled approach would be to incorporate the relevant physical constraints directly into the neural-network ansatz itself. We leave the development of such physically informed architectures to future work.


%
%
\section{Conclusion}
\label{sec:Conclusion}

In this work, we introduced \(\Sigma\)-Attention, a Transformer-based operator-learning framework designed to learn the self-energy of strongly correlated systems. By leveraging an encoder-only Transformer as an ansatz for approximating the self-energy operator, our approach combines complementary datasets from many-body perturbation theory, strong-coupling expansion, and exact diagonalization. This hybrid strategy allows \(\Sigma\)-Attention to learn an accurate approximation to the self-energy that remains valid across different parameter regimes. Our demonstration on the 1D Hubbard model at half-filling shows that \(\Sigma\)-Attention effectively captures the metal-to-insulator transition while addressing finite-size effects by extending predictions to larger system sizes. This framework not only harnesses the strengths of different established theoretical methods but also exploits the generalization capabilities of Transformer architectures to overcome their individual limitations.

Looking ahead, whether the proposed framework can generalize to non-half-filling cases remains an open question since the underlying physics becomes more complex. We anticipate that it is likely to be insufficient if only using training data from the half-filled regime; however, incorporating representative and accurate training data from non-half-filled regimes could potentially enable accurate predictions. This is an important direction that we leave for future investigation. More broadly, the promising results of this study open avenues for other research directions in the future. Potential extensions include: (i) incorporating additional physical parameters (e.g., temperature, chemical potential) and the corresponding numerical simulation data into the framework; (ii) extending the method to predict two-particle Green's functions and higher-order correlation functions; (iii) generalizing to 2D and 3D systems; and (iv) applying the methodology to real material systems. We believe that \(\Sigma\)-Attention offers a scalable and flexible pathway for advancing the study of strongly correlated materials and may pave the way for new insights into a variety of complex quantum phenomena.

\section{Acknowledgement}
 This material is based upon work supported by the U.S. Department of Energy, Office of Science, Office of Advanced Scientific Computing Research and Office of Basic Energy Sciences, Scientific Discovery through Advanced Computing (SciDAC) program under Award Number DE-SC0022198.  This work is also supported by the Center for Computational Study of Excited-State Phenomena in Energy Materials (C2SEPEM) at the Lawrence Berkeley National Laboratory, which is funded by the U.S. Department of Energy, Office of Science, Basic Energy Sciences, Materials Sciences and Engineering Division, under Contract No. DE-AC02-05CH11231, as part of the Computational Materials Sciences Program. This work is partially supported by the Simons Targeted Grant in Mathematics and Physical Sciences on Moire Materials Magic (Z. H.). This research used resources of the National Energy Research Scientific Computing Center, a DOE Office of Science User Facility supported by the Office of Science of the U.S. Department of Energy under Contract No. DE-AC02-05CH11231 using NERSC award BES ERCAP0029462 (project m4022) and ASCR-ERCAP m1027. The Flatiron Institute is a division of the Simons Foundation. Y. Zhu gratefully acknowledges the invaluable discussions with colleagues and friends, including Lin Lin, Efekan Kokcu, Lei Zhang, Harish Bhat, Sergei Iskakov, Emanuel Gull, Vojtech Vlcek, Jason Kaye, Harrison LaBollita, Zhouquan Wan and Peizhi Mai. 

\section{Data Availability Statement}
The data that support the findings of this study are available from the authors upon reasonable request.



%
%
\appendix
\section{Numerical simulation details}\label{sec:details}

\subsection{Training Data}
\label{subsec:data_source}

MBPT, SCE, and ED are employed to generate the combined training dataset for \(\Sigma\)-attention. The MBPT data are obtained from the equilibrium Matsubara Dyson's equation solver in the \texttt{NESSi} Package \cite{schuler2020nessi}, using the second-order Born (2ndB) self-energy approximation for the Hubbard model. The SCE results are based on the analytical expression for the Green's function of the 1D Hubbard model derived by Pairault et al. \cite{pairault1998strong}:

\begin{widetext}
\begin{equation*}
G^{-1}(k,\mathrm i\omega_n)= 2t\cos(k) + \Biggl\{
\frac{\mathrm i\omega_n}{(\mathrm i\omega_n)^2-\frac{U^2}{4}}
+\frac{6t^2U^2\,\mathrm i\omega_n}{4\left[(\mathrm i\omega_n)^2-\frac{U^2}{4}\right]^3}
+6t^3\cos(k)\Biggl[
\frac{(\beta U/4)\tanh(\beta U/4)}
{\left[(\mathrm i\omega_n)^2-\frac{U^2}{4}\right]^2}
+\frac{U^2\Bigl(2(\mathrm i\omega_n)^2-\frac{U^2}{4}\Bigr)}
{4\left[(\mathrm i\omega_n)^2-\frac{U^2}{4}\right]^4}
\Biggr]
\Biggr\}^{-1}.
\end{equation*}
\end{widetext}
The corresponding self-energy is obtained by solving Dyson's equation \eqref{dyson_eqn_inv} using the above $G(k,\mathrm i\omega_n)$ and the Hartree-Fock (HF) solution as $G_0(k,\mathrm i\omega_n)$. The \texttt{EDLib} \cite{iskakov2018exact} is the solver we used to generate the ED training data. Note that for the Hubbard model with $\beta=20$, setting the eigenvalue number threshold to be $N=20$ would include sufficient exited states for getting an accurate enough Green's function. The corresponding self-energy is also obtained by solving Dyson's equation \eqref{dyson_eqn_inv}. After finishing all these calculation, we frequently use \texttt{pydlr} \cite{kaye2022libdlr,kaye2022discrete} to accurately interpolate discretized functions and perform transformations such as $G_{ij}(\tau)\leftrightarrow G_{ij}(\mathrm i\omega_n) \leftrightarrow G(k,\mathrm i\omega_n)\leftrightarrow G(k,\tau)$ for both the Green's function and the self-energy.

As illustrated in FIG. \ref{fig:data_structure}, in the overlapping regime, MBPT and SCE generated data are cross-checked with the ED result. Through this comparison, we determined that for the 1D Hubbard model \eqref{hubbard} and $N_s\geq 32$, the MBPT result is valid for $U\leq 1.5$ ($l_{\infty}$-error for $G(k,\tau)$ is $\sim 10^{-3}$). The SCE is valid for $U\geq 6.0$ ($l_{\infty}$-error for $G(k,\tau)$ is $\sim 10^{-3}$).
\subsection{Learning and Training details}
\label{subsec:Learning and Training details}
\(\Sigma\)-attention takes as input the Hartree-Fock Green's function \(G_0(k,\tau)\) (or equivalently \(G_0(k,\mathrm i\omega_n)\)) along with the Hubbard interaction \(U\) and outputs the predicted self-energy \(\Sigma(k,\tau)\) (or \(\Sigma(k,\mathrm i\omega_n)\)). The entire neural network is implemented in PyTorch and trained on NERSC GPUs. Because the main functionalities of PyTorch such as auto-differentiations only support real-valued data, we adopt the \(\tau\)-space formulation, i.e.,
\[
G_0(k,\tau), U \xrightarrow{\text{NN}} \Sigma(k,\tau).
\]
For an inverse temperature \(\beta = 20\), we discretize \(G_0(k,\tau)\) using an equally spaced \(\tau\)-grid \(\{\tau_i\}_{i=1}^{N_\tau} \subset [0,\beta]\) with \(N_\tau = 101\). The momentum \(k\)-points provide a natural positional encoding via \(\cos(k_j)\), where
\[
k_j = \frac{2\pi j}{N_s},\quad j = -\frac{N_s}{2},\ldots,\frac{N_s}{2},
\]
and \(N_s\) is the total number of sites. The HF Green's function \(G_0(k,\tau_i)\), the scalar \(U\) (broadcasted as an all-ones vector \(\vec{J}\) of dimension \(N_s\times1\)), and the positional encoding \(\cos(k_j)\) are concatenated along the columns to form an input matrix:
\begin{align*}
\{G_0(k_j,\tau_i), U\vec{J},\cos(k_j)\}\xrightarrow{\text{Concat}} \text{InputMat}_{N_s\times(N_{\tau}+2)}
\end{align*}
where the feature dimension $n_f= N_{\tau}+2=103$. Our encoder-only Transformer takes in InputMat and processes it through 2 identical {\em Multi-Head Attention} layers and one final projection layer to generate the final output: self-energy $\Sigma(k_i,\tau_j)\in \mathbb{R}^{N_s\times 101}$. Conceptually, the Multi-Head Attention is just the concatenation of the output of single-head Attention along the columns, which is projected back to the target output using a new matrix $W^O$. The corresponding computational rule therefore is
\cite{vaswani2017attention}:
\begin{align*}
\text{MultiHead}(\text{InputMat}) &=\text{Concat}(\text{Head}_1,\cdots,\text{Head}_{16})W^O\\
\text{Head}_i &= \text{Attention}[\tilde  Q_i,\tilde K_i,\tilde V_i]
\end{align*}
where 
\begin{align*}
\tilde Q_i &=\text{InputMat}W^{Q}_i,\\
\tilde K_i &=\text{InputMat}W^{K}_i,\\
\tilde V_i &=\text{InputMat}W^{V}_i.
\end{align*}
In our setting, each Multi-Head Attention has 16 heads and the hidden dimension of each head is \(n_d = 16\). Consequently, the coefficient matrices \(W_i^Q\), \(W_i^K\), and \(W_i^V\) in each head are of dimension \(\mathbb{R}^{103 \times 16}\) (since the input feature dimension $n_f=103$). The Multi-head Attention output projection matrix \(W^O\) has dimension \(\mathbb{R}^{256 \times 103}\), mapping the concatenated 256 head dimensions back to 103. Finally, a linear output layer with weight matrix \(O \in \mathbb{R}^{103 \times 101}\) transforms the hidden representation of dimension \(\mathbb{R}^{N_s \times 103}\) into the final output \(\Sigma(k_i,\tau_j) \in \mathbb{R}^{N_s \times 101}\).

\begin{figure}
\centering
\includegraphics[width=7cm]{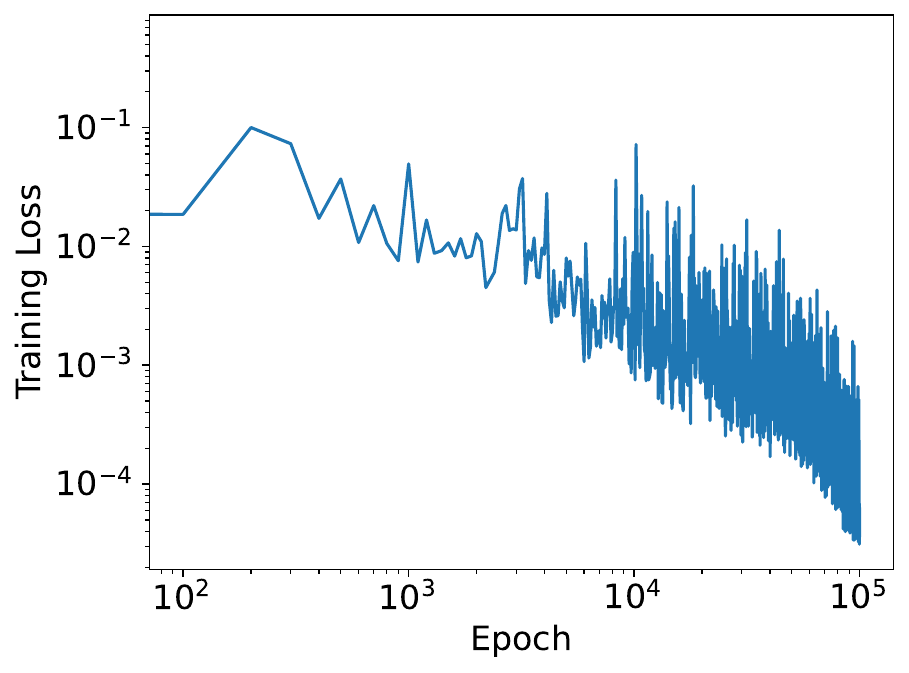}
\caption{Training loss decays for a representative run of $\Sigma$-Attention. The large error fluctuation is due to the randomized batch sweeping.}
\label{fig:loss_decay}
\end{figure}

Note that the weighted inner product in self-Attention has some similarities with the 2ndB (bare) self-energy:
\begin{align}\label{2ndB}
\Sigma^{2ndB}(k,\tau) &= \frac{U^2}{N_k^2}\sum_{pq}G_0(k-q,\tau)G_0(p+q,\tau)G_0(p,-\tau)  
\end{align}
$\Sigma$-Attention is trained in a supervised way by matching the NN-predicted self-energy with the ground-truth values obtained via MBPT, SCE, and ED within their respective ranges of applicability. Because the network is optimized on a diverse dataset \(\{G_0, U\}\), the corresponding self-energy \(\Sigma[U,G_0]\) exhibits varying magnitudes across different \(U\) values. This variability can cause optimization difficulties, as the error may fluctuate and eventually favor data points with larger \(U\), leading to underfitting in the small-\(U\) regime. To address this and to ensure that the NN-predicted self-energy remains positive (and hence physically meaningful), we rescale the target self-energy according to
\[
\Sigma_S[U,G_0](k,\tau) = \frac{\sqrt{|\Sigma(k,\tau)|}}{0.13\, U^2}.
\]
where the constant $0.13$ is obtained by numerical fitting to ensure the scaled $\Sigma_S[U,G_0](k,\tau)$ for different $U$-values will have the same magnitude as of $G_0(k,\tau)$, which will make the optimization easier. Thus, the direct output of \(\Sigma\)-attention is \(\Sigma_S[U,G_0](k,\tau)\). The loss function used during optimization is a combined \(l_1 + l_2\) loss, given by
\begin{align*}
\mathcal{L} &= \sum_{ij}\left|\Sigma_S^{NN}[U,G_0](k_j,\tau_i) - \Sigma_S^{\text{target}}[U,G_0](k_j,\tau_i)\right|\\
&+\sum_{ij}\left|\Sigma_S^{NN}[U,G_0](k_j,\tau_i) - \Sigma_S^{\text{target}}[U,G_0](k_j,\tau_i)\right|^2.
\end{align*}
where $\Sigma^{NN}_S(k_j,\tau_j)$ is the output of $\Sigma$-Attention that corresponds to the input $G_0,U$. During training, given the large combined dataset, we employ the randomized batch sweeping technique \cite{zhu2023learning}. Namely, in each epoch, we randomly sample a single (or a small batch) input/output pair:
\(\{\text{InputMat},\, \Sigma_S[U,G_0](k_j,\tau_i)\}\),
i.e., a data point from the entire dataset, and perform optimization. For a sufficient amount of training epochs, the randomized training almost surely ensures that every data point from the dataset is visited. This approach shares the same spirit of the stochastic gradient descent and is particularly useful for operator-learning tasks \cite{zhu2023learning,bassi2024learning,zhu2025predicting}. The training epochs for examples in Section \ref{sec:app} are set to be $N=100000$, Adam optimizer with adaptive learning rate is used to optimize the NN. In PyTorch, this can be done simply using the function \texttt{torch.optim.lr\_scheduler.CosineAnnealingLR}. FIG~\ref{fig:loss_decay} illustrates the decay of the training loss for a representative run.

\section{Additional numerical results}
\subsection{Hubbard model of different sizes}\label{app:add_numerial}
\begin{figure*}
\centering
\includegraphics[width=17cm]{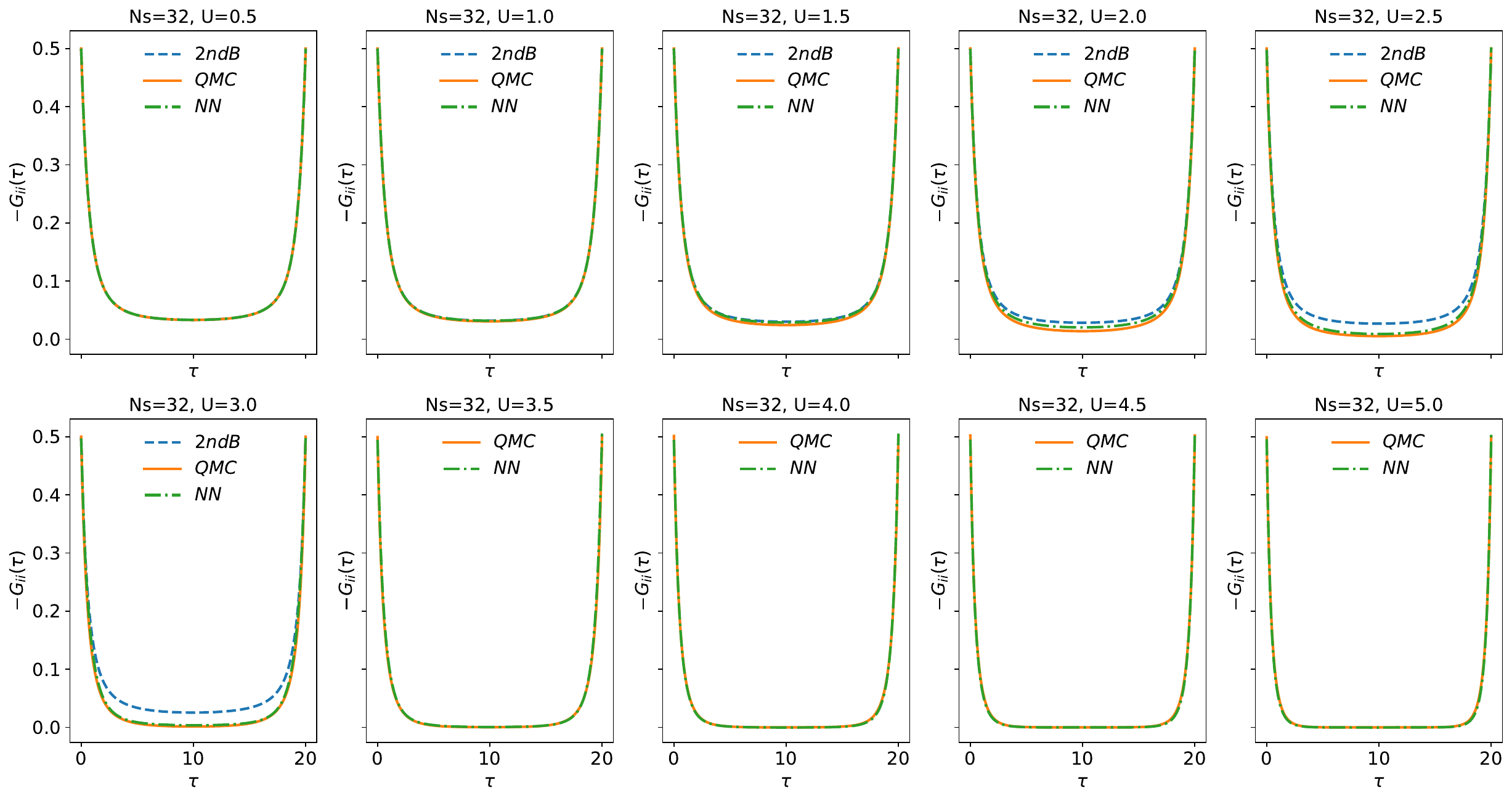}
\caption{Prediction of onsite Green's function $G_{ii}(\tau)$ for Hubbard model with $N_s=32$.}
\label{fig:Gii_N32}
\end{figure*}

\begin{figure*}
\centering
\includegraphics[width=17cm]{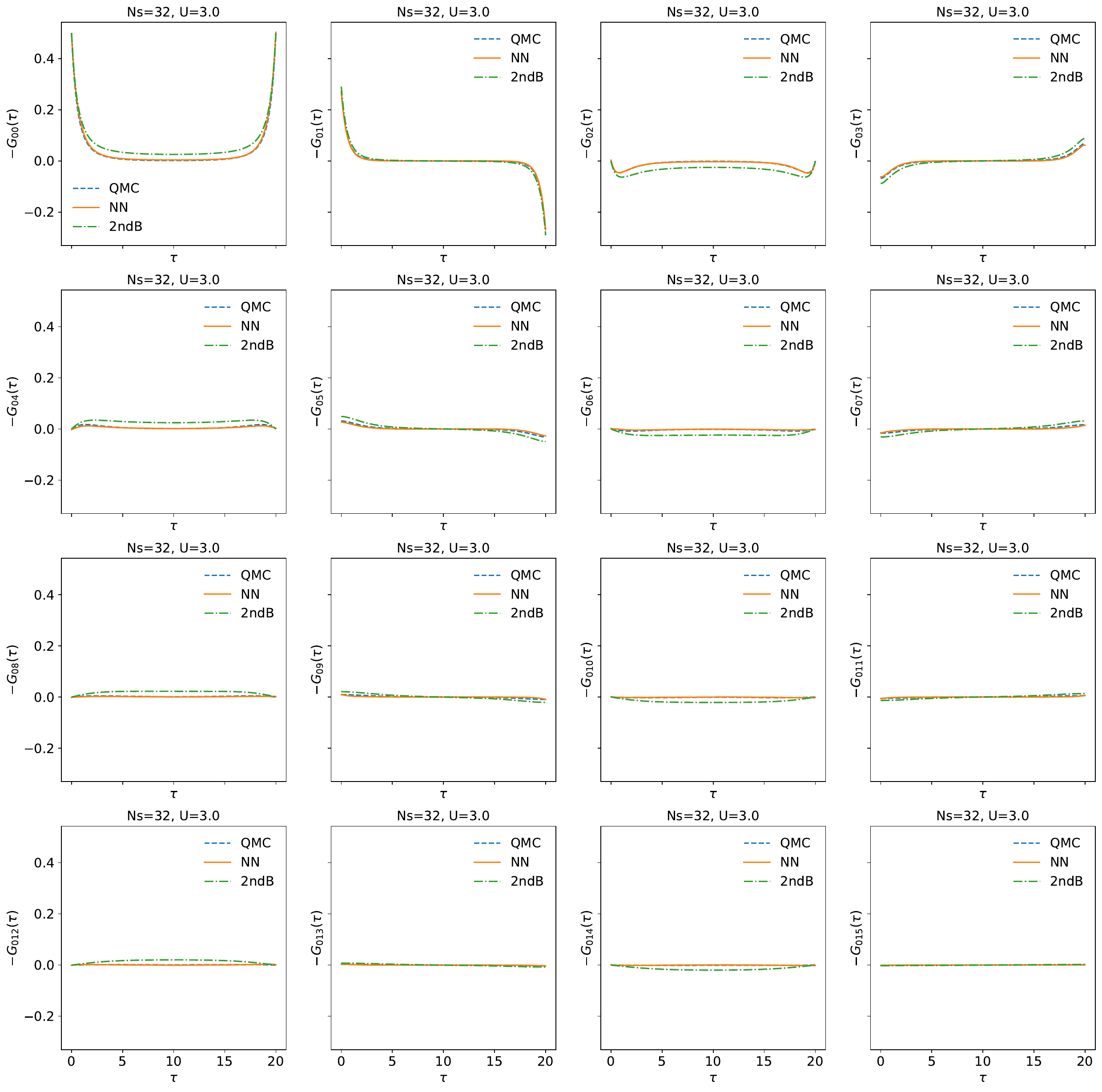}
\caption{Prediction of the off-site Green's function $G_{0j}(\tau)$ for Hubbard model with $N_s=32$. The data shown corresponds to an interaction strength of $U=3.0$. Similar trends are observed for other values of $U$. }
\label{fig:Gij_N32}
\end{figure*}

\begin{figure*}
\centering
\includegraphics[width=17cm]{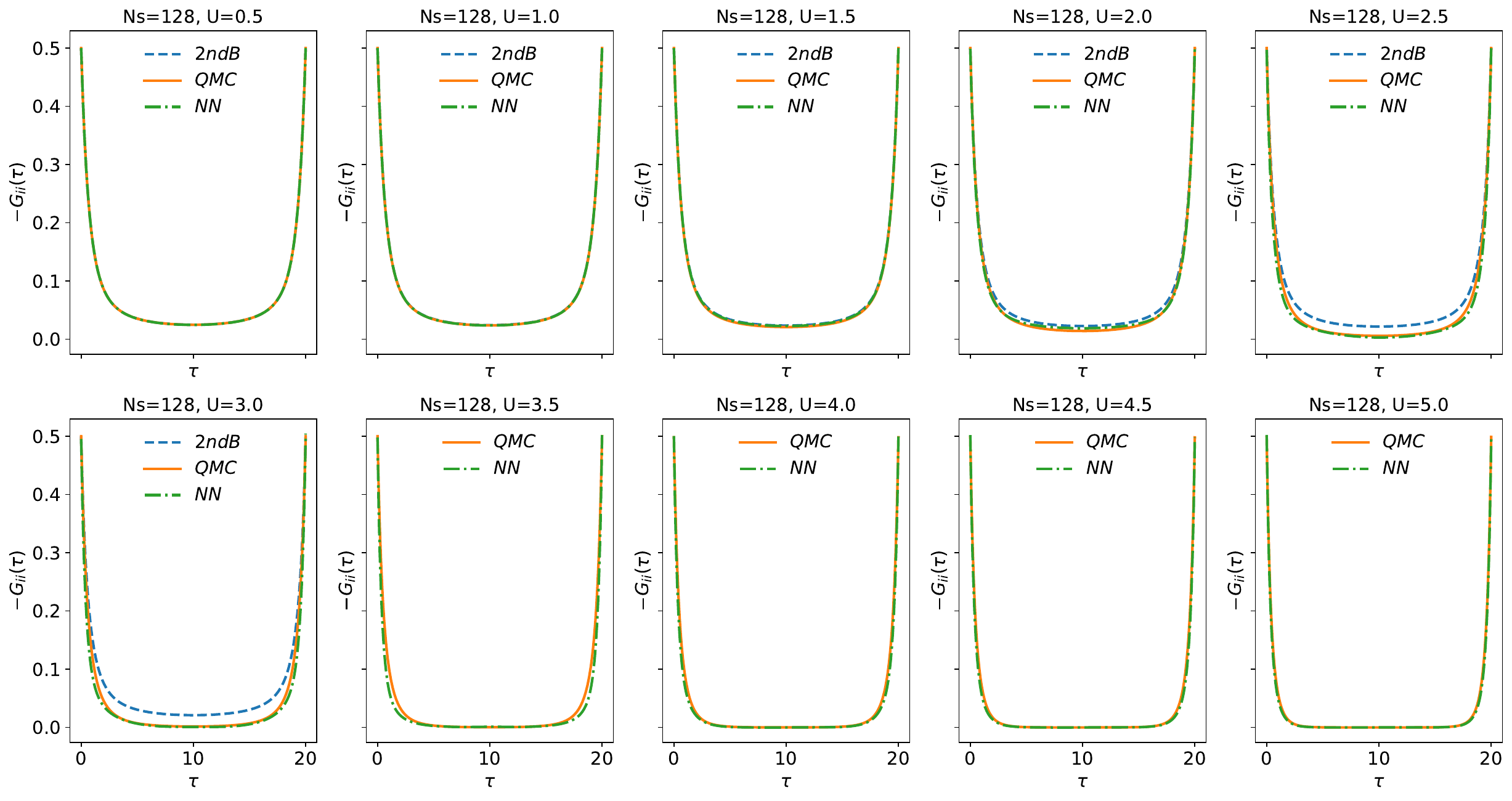}
\caption{Prediction of onsite Green's function $G_{ii}(\tau)$ for Hubbard model with $N_s=128$.}
\label{fig:Gii_N128}
\end{figure*}

\begin{figure*}
\centering
\includegraphics[width=17cm]{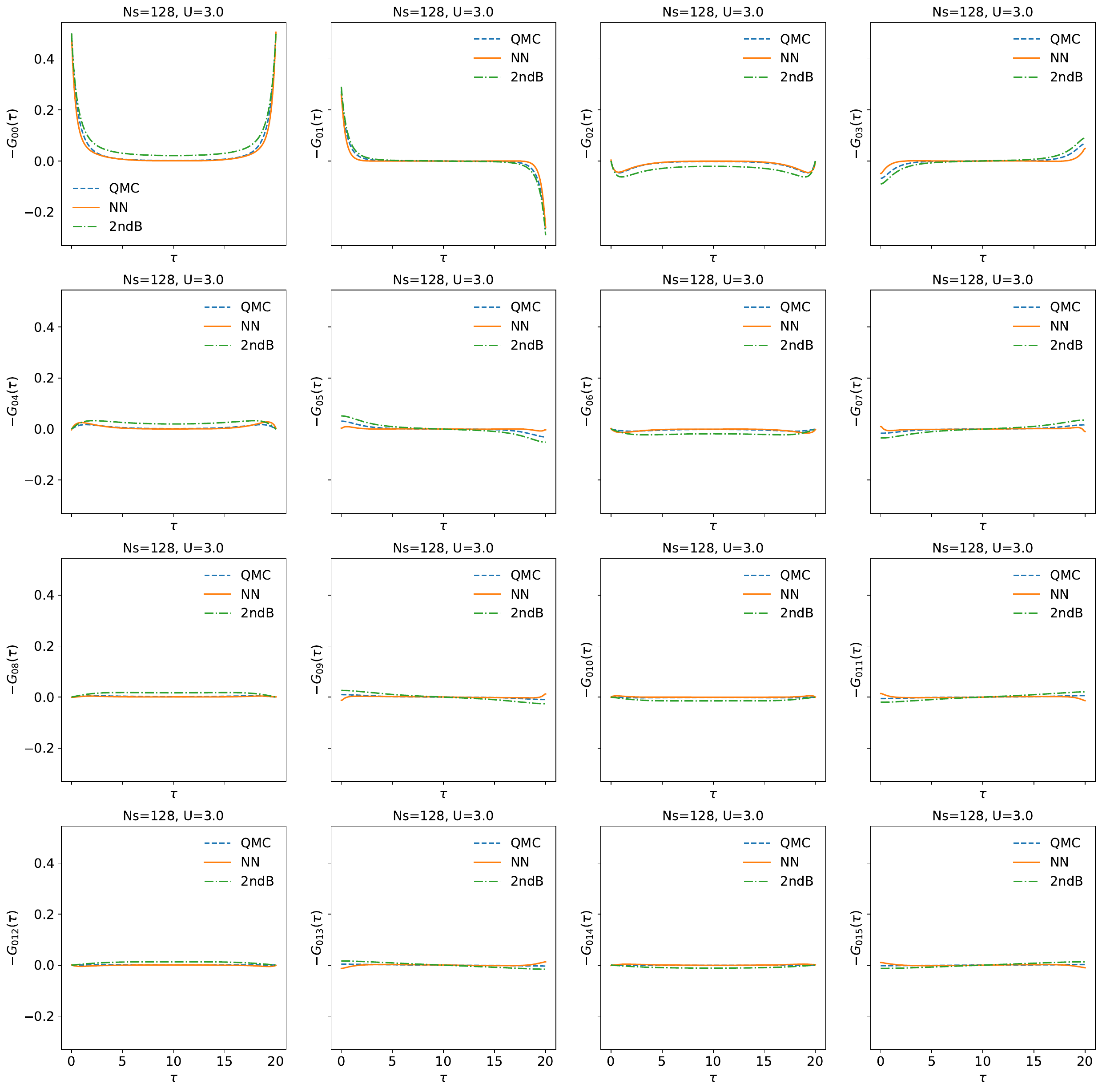}
\caption{Prediction of the off-site Green's function $G_{0j}(\tau)$ for Hubbard model with $N_s=128$. The data shown corresponds to an interaction strength of $U=3.0$. Similar trends are observed for other values of $U$. }
\label{fig:Gij_N128}
\end{figure*}

In this section, we present additional numerical results obtained by $\Sigma$-Attention for Hubbard model with system sizes $N_s=32$ and $N_s=128$. Figure~\ref{fig:Gii_N32} displays the approximated onsite Green's function $G_{ii}(\tau)$. Figure~\ref{fig:Gij_N32} illustrates the first few off-site Green's functions $G_{0j}(\tau)$. Figures~\ref{fig:Gii_N128}--\ref{fig:Gij_N128} present analogous results for the system size $N_s=128$. The information we obtained from these figures is consistent with the main conclusions presented in the paper.

\subsection{Effect of causal projection}
\begin{figure}
\centering
\includegraphics[width=7cm]{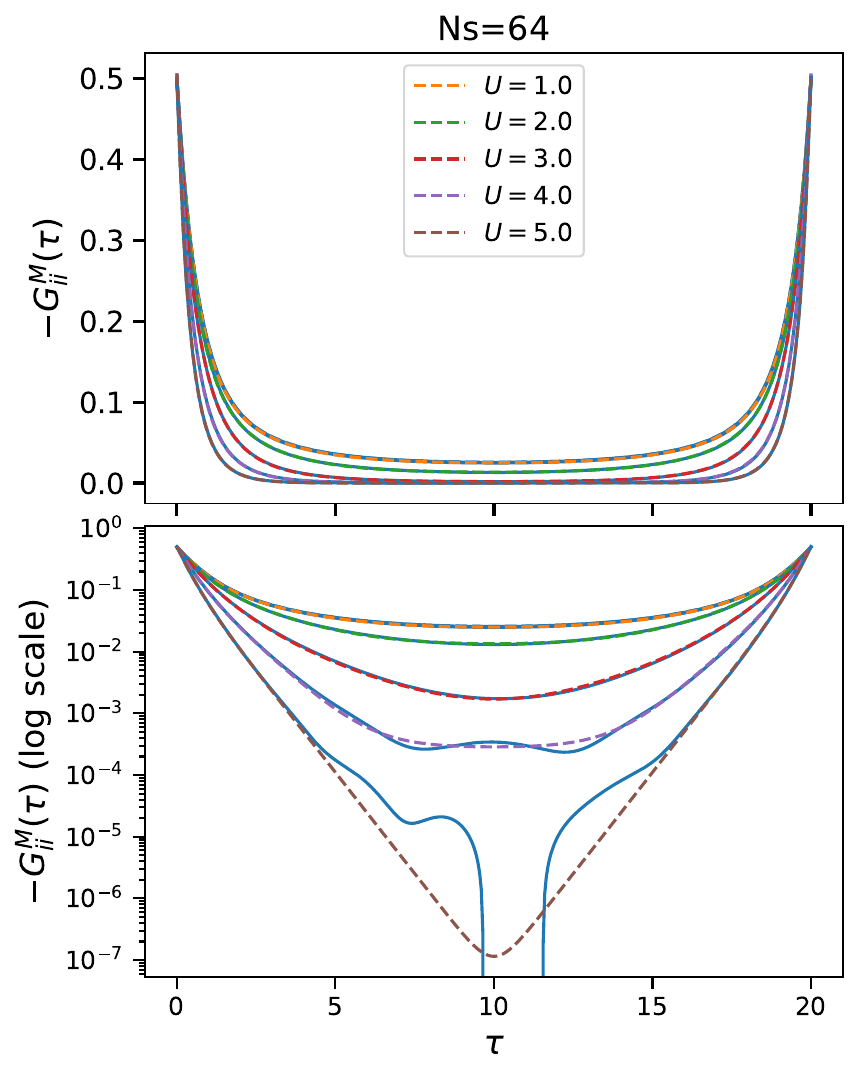}
\caption{
    Comparison of NN-predicted Green's function $G_{ii}(\tau)$ before and after the causal projection, where the solid line represents the original result and the dashed line represents the projected result. The data is displayed both on the regular (top) and log (bottom) scale.}
\label{fig:Gii_filter_vs_unfilter}
\end{figure}

In the main text, we apply the adaptive pole-fitting techniques developed in~\cite{huang2023robust} to both the AFQMC and the NN-predicted Green's function to ensure its causality. Figure~\ref{fig:Gii_filter_vs_unfilter} compares the NN-predicted $G_{ii}(\tau)$ before and after the causal projection is applied. Quantitatively, we can see that the causal projection does not change the function significantly but effectively filters out numerical noise and ensures that the resulting Green’s function satisfies the required causality conditions. This is done by doing discrete Lehmann representation fitting for the Matsubara Green's function data~\cite{huang2023robust}.
\subsection{Importance of combined training datasets}\label{app:importance_dataset}
\begin{figure*}
\centering
\includegraphics[width=5.3cm]{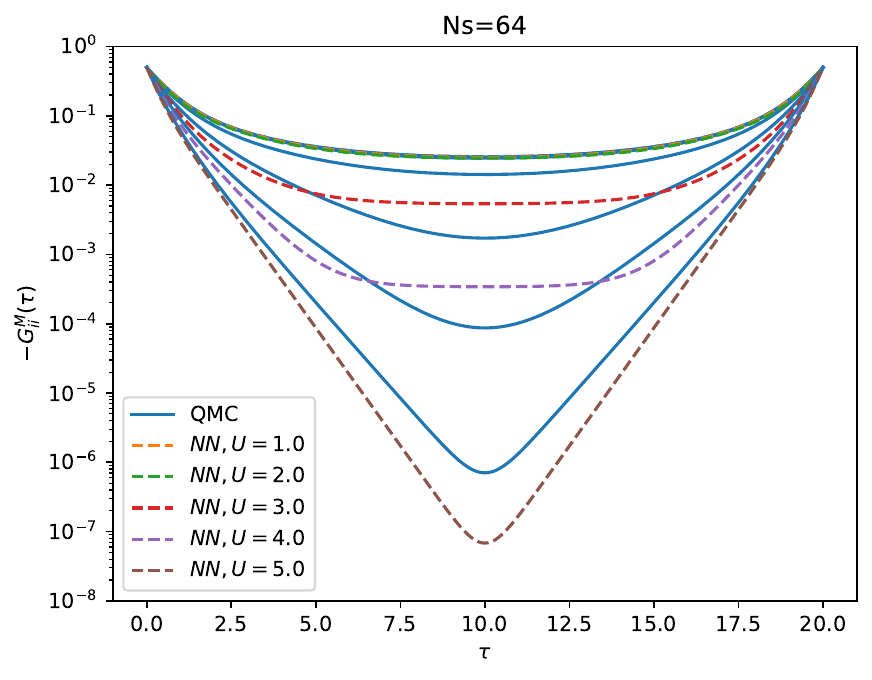}
\includegraphics[width=5.3cm]{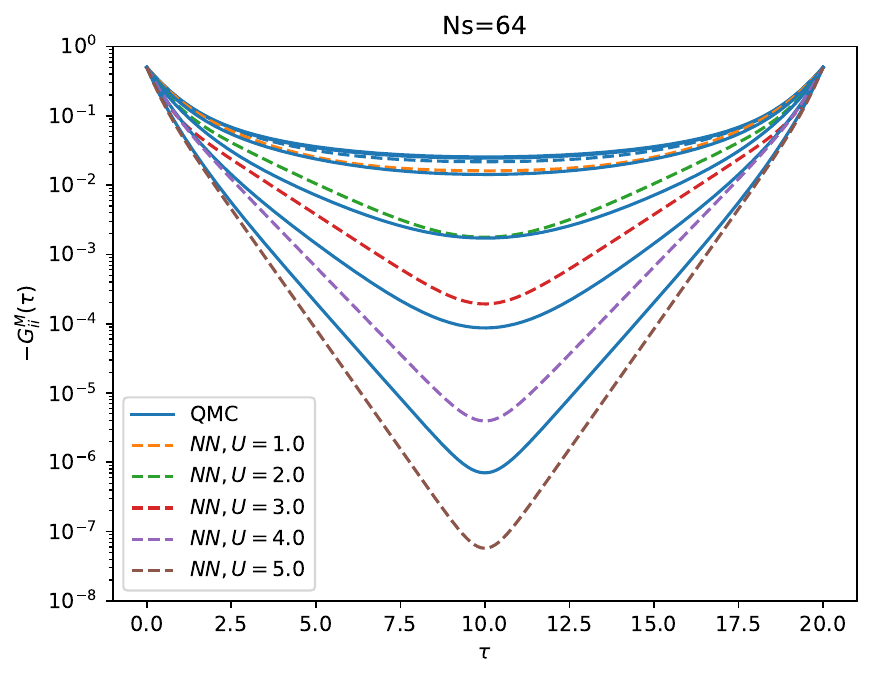}
\includegraphics[width=5.3cm]{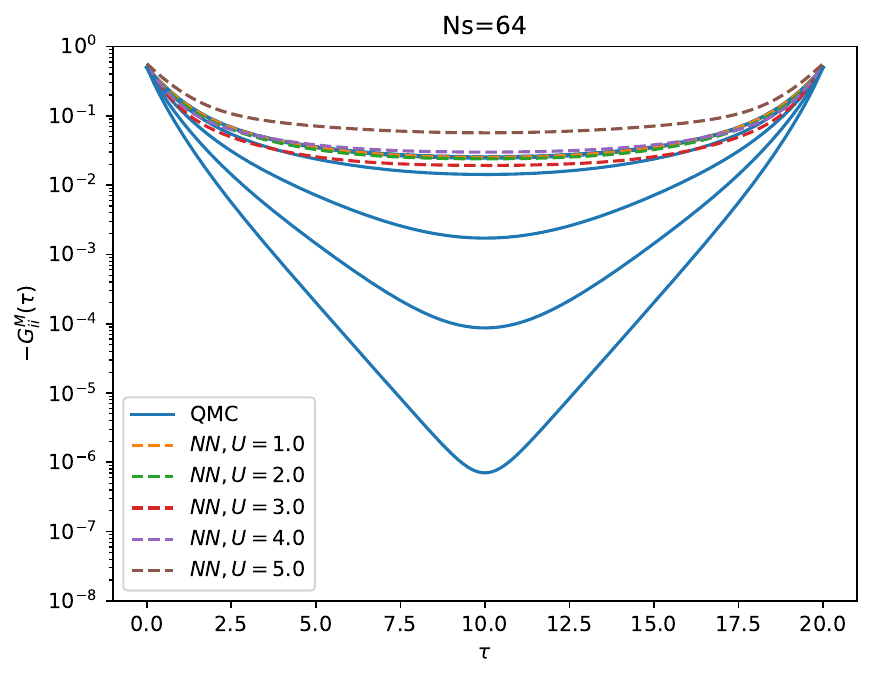}
\caption{Comparison of the onsite Green's function $G_{ii}(\tau)$ computed using different combinations of the training datasets. Specifically, they are: (Left) MBPT+SCE (Middle) ED+SCE (Right) ED+MBPT}
\label{fig:incomplete_data}
\end{figure*}

To demonstrate the importance of each component in the combined MBPT+ED+SCE dataset and show each of them influences the final outcome of $\Sigma$-Attention. Figure~\ref{fig:incomplete_data} presents the NN prediction for $N_s=64$ when one component is excluded from the training dataset. Compared with Figures~\ref{fig:Gii_log}, one notices that in all cases the prediction error increases. In particular, excluding the SCE data results in completely wrong predictions (the metal-to-insulator transition is not captured), clearly underscoring its importance in the learning process. This is as expected since the SCE result informs $\Sigma$-Attention of the insulting state as $U\rightarrow+\infty$.

%
%
    

\bibliography{apssamp}

@article{kaufmann2023ana_cont,
  title={ana\_cont: Python package for analytic continuation},
  author={Kaufmann, Josef and Held, Karsten},
  journal={Computer Physics Communications},
  volume={282},
  pages={108519},
  year={2023},
  publisher={Elsevier}
}

@article{nelson2019machine,
  title={Machine learning density functional theory for the Hubbard model},
  author={Nelson, James and Tiwari, Rajarshi and Sanvito, Stefano},
  journal={Physical Review B},
  volume={99},
  number={7},
  pages={075132},
  year={2019},
  publisher={APS}
}

@article{huang2023robust,
  title={Robust analytic continuation of Green's functions via projection, pole estimation, and semidefinite relaxation},
  author={Huang, Zhen and Gull, Emanuel and Lin, Lin},
  journal={Physical Review B},
  volume={107},
  number={7},
  pages={075151},
  year={2023},
  publisher={APS}
}

@article{pairault1998strong,
  title={Strong-coupling expansion for the Hubbard model},
  author={Pairault, St{\'e}phane and S{\'e}n{\'e}chal, David and Tremblay, A-MS},
  journal={Physical review letters},
  volume={80},
  number={24},
  pages={5389},
  year={1998},
  publisher={American Physical Society}
}

@article{pairault2000strong,
  title={Strong-coupling perturbation theory of the Hubbard model},
  author={Pairault, Stephane and Senechal, David and Tremblay, A-MS},
  journal={The European Physical Journal B-Condensed Matter and Complex Systems},
  volume={16},
  pages={85--105},
  year={2000},
  publisher={Springer}
}

@article{stanescu2004strong,
  title={Strong coupling theory for interacting lattice models},
  author={Stanescu, Tudor D and Kotliar, Gabriel},
  journal={Physical Review B—Condensed Matter and Materials Physics},
  volume={70},
  number={20},
  pages={205112},
  year={2004},
  publisher={APS}
}

@article{metzner1991linked,
  title={Linked-cluster expansion around the atomic limit of the Hubbard model},
  author={Metzner, Walter},
  journal={Physical Review B},
  volume={43},
  number={10},
  pages={8549},
  year={1991},
  publisher={American Physical Society}
}

@article{dupuis2000strong,
  title={A strong-coupling expansion for the Hubbard model},
  author={Dupuis, N and Pairault, S},
  journal={International Journal of Modern Physics B},
  volume={14},
  number={24},
  pages={2529--2560},
  year={2000},
  publisher={World Scientific}
}

@article{qin2016benchmark,
  title={Benchmark study of the two-dimensional Hubbard model with auxiliary-field quantum Monte Carlo method},
  author={Qin, Mingpu and Shi, Hao and Zhang, Shiwei},
  journal={Physical Review B},
  volume={94},
  number={8},
  pages={085103},
  year={2016},
  publisher={APS}
}

@article{zhang1997constrained,
  title={Constrained path Monte Carlo method for fermion ground states},
  author={Zhang, Shiwei and Carlson, Joseph and Gubernatis, James E},
  journal={Physical Review B},
  volume={55},
  number={12},
  pages={7464},
  year={1997},
  publisher={APS}
}

@article{zhang2003quantum,
  title={Quantum Monte Carlo method using phase-free random walks with Slater determinants},
  author={Zhang, Shiwei and Krakauer, Henry},
  journal={Physical review letters},
  volume={90},
  number={13},
  pages={136401},
  year={2003},
  publisher={APS}
}

@article{gull2011continuous,
  title={Continuous-time Monte Carlo methods for quantum impurity models},
  author={Gull, Emanuel and Millis, Andrew J and Lichtenstein, Alexander I and Rubtsov, Alexey N and Troyer, Matthias and Werner, Philipp},
  journal={Reviews of Modern Physics},
  volume={83},
  number={2},
  pages={349--404},
  year={2011},
  publisher={APS}
}

@article{schollwock2005density,
  title={The density-matrix renormalization group},
  author={Schollw{\"o}ck, Ulrich},
  journal={Reviews of modern physics},
  volume={77},
  number={1},
  pages={259--315},
  year={2005},
  publisher={APS}
}

@article{white1998strongly,
  title={Strongly correlated electron systems and the density matrix renormalization group},
  author={White, Steven R},
  journal={Physics Reports},
  volume={301},
  number={1-3},
  pages={187--204},
  year={1998},
  publisher={Elsevier}
}

@book{fulde1995electron,
  title={Electron correlations in molecules and solids},
  author={Fulde, Peter},
  volume={100},
  year={1995},
  publisher={Springer Science \& Business Media}
}

@article{senechal2002cluster,
  title={Cluster perturbation theory for Hubbard models},
  author={S{\'e}n{\'e}chal, David and Perez, Danny and Plouffe, Dany},
  journal={Physical Review B},
  volume={66},
  number={7},
  pages={075129},
  year={2002},
  publisher={APS}
}

@article{senechal2000spectral,
  title={Spectral weight of the Hubbard model through cluster perturbation theory},
  author={S{\'e}n{\'e}chal, D and Perez, Danny and Pioro-Ladriere, M},
  journal={Physical review letters},
  volume={84},
  number={3},
  pages={522},
  year={2000},
  publisher={APS}
}

@article{bartkowiak1995high,
  title={High-temperature series expansion for the extended Hubbard model},
  author={Bartkowiak, M and Henderson, JA and Oitmaa, J and De Brito, PE},
  journal={Physical Review B},
  volume={51},
  number={20},
  pages={14077},
  year={1995},
  publisher={APS}
}

@article{perepelitsky2016transport,
  title={Transport and optical conductivity in the Hubbard model: A high-temperature expansion perspective},
  author={Perepelitsky, Edward and Galatas, Andrew and Mravlje, Jernej and {\v{Z}}itko, Rok and Khatami, Ehsan and Shastry, B Sriram and Georges, Antoine},
  journal={Physical Review B},
  volume={94},
  number={23},
  pages={235115},
  year={2016},
  publisher={APS}
}

@article{iskakov2018exact,
  title={Exact diagonalization library for quantum electron models},
  author={Iskakov, Sergei and Danilov, Michael},
  journal={Computer Physics Communications},
  volume={225},
  pages={128--139},
  year={2018},
  publisher={Elsevier}
}

@article{potthoff2004non,
  title={Non-perturbative construction of the {L}uttinger-{W}ard functional},
  author={Potthoff, Michael},
  journal={Condens. Mat. Phys.},
  volume ={9},
  page={557},
  year={2006}
}

@article{schuler2020nessi,
  title={NESSi: The non-equilibrium systems simulation package},
  author={Sch{\"u}ler, Michael and Gole{\v{z}}, Denis and Murakami, Yuta and Bittner, Nikolaj and Herrmann, Andreas and Strand, Hugo UR and Werner, Philipp and Eckstein, Martin},
  journal={Computer Physics Communications},
  volume={257},
  pages={107484},
  year={2020},
  publisher={Elsevier}
}

@book{stefanucci2013nonequilibrium,
  title={Nonequilibrium Many-Body Theory of Quantum Systems: A Modern Introduction},
  author={Stefanucci, G. and van Leeuwen, R.},
  isbn={9780521766173},
  lccn={2012050475},
  year={2013},
  publisher={Cambridge University Press}
}

@article{zhu2023learning,
  title={Learning stochastic dynamics with statistics-informed neural network},
  author={Zhu, Yuanran and Tang, Yu-Hang and Kim, Changho},
  journal={Journal of Computational Physics},
  volume={474},
  pages={111819},
  year={2023},
  publisher={Elsevier}
}

@article{bassi2024learning,
  title={Learning nonlinear integral operators via recurrent neural networks and its application in solving integro-differential equations},
  author={Bassi, Hardeep and Zhu, Yuanran and Liang, Senwei and Yin, Jia and Reeves, Cian C and Vl{\v{c}}ek, Vojt{\v{e}}ch and Yang, Chao},
  journal={Machine Learning with Applications},
  volume={15},
  pages={100524},
  year={2024},
  publisher={Elsevier}
}

@article{georges1996dynamical,
  title={Dynamical mean-field theory of strongly correlated fermion systems and the limit of infinite dimensions},
  author={Georges, Antoine and Kotliar, Gabriel and Krauth, Werner and Rozenberg, Marcelo J},
  journal={Reviews of modern physics},
  volume={68},
  number={1},
  pages={13},
  year={1996},
  publisher={APS}
}

@article{kotliar2001cellular,
  title={Cellular dynamical mean field approach to strongly correlated systems},
  author={Kotliar, Gabriel and Savrasov, Sergej Y and P{\'a}lsson, Gunnar and Biroli, Giulio},
  journal={Physical review letters},
  volume={87},
  number={18},
  pages={186401},
  year={2001},
  publisher={APS}
}

@article{hettler2000dynamical,
  title={Dynamical cluster approximation: Nonlocal dynamics of correlated electron systems},
  author={Hettler, Mattias H and Mukherjee, M and Jarrell, Mark and Krishnamurthy, Hulikal R},
  journal={Physical Review B},
  volume={61},
  number={19},
  pages={12739},
  year={2000},
  publisher={APS}
}

@article{macridin2006phase,
  title={Phase separation in the Hubbard model using the dynamical cluster approximation},
  author={Macridin, Alexandru and Jarrell, Mark and Maier, Th},
  journal={Physical Review B—Condensed Matter and Materials Physics},
  volume={74},
  number={8},
  pages={085104},
  year={2006},
  publisher={APS}
}

@article{kaye2022libdlr,
  title={libdlr: Efficient imaginary time calculations using the discrete Lehmann representation},
  author={Kaye, Jason and Chen, Kun and Strand, Hugo UR},
  journal={Computer Physics Communications},
  volume={280},
  pages={108458},
  year={2022},
  publisher={Elsevier}
}

@article{kaye2022discrete,
  title={Discrete Lehmann representation of imaginary time Green's functions},
  author={Kaye, Jason and Chen, Kun and Parcollet, Olivier},
  journal={Physical Review B},
  volume={105},
  number={23},
  pages={235115},
  year={2022},
  publisher={APS}
}

@article{kotliar2006electronic,
  title={Electronic structure calculations with dynamical mean-field theory},
  author={Kotliar, Gabriel and Savrasov, Sergej Y and Haule, Kristjan and Oudovenko, Viktor S and Parcollet, O and Marianetti, CA},
  journal={Reviews of Modern Physics},
  volume={78},
  number={3},
  pages={865--951},
  year={2006},
  publisher={APS}
}

@article{van2010diagrammatic,
  title={Diagrammatic monte carlo},
  author={Van Houcke, Kris and Kozik, Evgeny and Prokof’ev, Nikolay and Svistunov, Boris},
  journal={Physics Procedia},
  volume={6},
  pages={95--105},
  year={2010},
  publisher={Elsevier}
}

@article{rossi2017determinant,
  title={Determinant diagrammatic Monte Carlo algorithm in the thermodynamic limit},
  author={Rossi, Riccardo},
  journal={Physical review letters},
  volume={119},
  number={4},
  pages={045701},
  year={2017},
  publisher={APS}
}

@article{liu2025accurate,
  title={Accurate Simulation of the Hubbard Model with Finite Fermionic Projected Entangled Pair States},
  author={Liu, Wen-Yuan and Zhai, Huanchen and Peng, Ruojing and Gu, Zheng-Cheng and Chan, Garnet Kin},
  journal={arXiv preprint arXiv:2502.13454},
  year={2025}
}

@article{chan2011density,
  title={The density matrix renormalization group in quantum chemistry},
  author={Chan, Garnet Kin-Lic and Sharma, Sandeep},
  journal={Annual review of physical chemistry},
  volume={62},
  number={1},
  pages={465--481},
  year={2011},
  publisher={Annual Reviews}
}

@article{chan2016matrix,
  title={Matrix product operators, matrix product states, and ab initio density matrix renormalization group algorithms},
  author={Chan, Garnet Kin and Keselman, Anna and Nakatani, Naoki and Li, Zhendong and White, Steven R},
  journal={The Journal of chemical physics},
  volume={145},
  number={1},
  year={2016},
  publisher={AIP Publishing}
}

@article{ainslie2020etc,
  title={ETC: Encoding long and structured inputs in transformers},
  author={Ainslie, Joshua and Ontanon, Santiago and Alberti, Chris and Cvicek, Vaclav and Fisher, Zachary and Pham, Philip and Ravula, Anirudh and Sanghai, Sumit and Wang, Qifan and Yang, Li},
  journal={arXiv preprint arXiv:2004.08483},
  year={2020}
}

@article{lu2024deepseek,
  title={Deepseek-vl: towards real-world vision-language understanding},
  author={Lu, Haoyu and Liu, Wen and Zhang, Bo and Wang, Bingxuan and Dong, Kai and Liu, Bo and Sun, Jingxiang and Ren, Tongzheng and Li, Zhuoshu and Yang, Hao and others},
  journal={arXiv preprint arXiv:2403.05525},
  year={2024}
}

@article{kissas2022learning,
  title={Learning operators with coupled attention},
  author={Kissas, Georgios and Seidman, Jacob H and Guilhoto, Leonardo Ferreira and Preciado, Victor M and Pappas, George J and Perdikaris, Paris},
  journal={Journal of Machine Learning Research},
  volume={23},
  number={215},
  pages={1--63},
  year={2022}
}

@article{calvello2024continuum,
  title={Continuum attention for neural operators},
  author={Calvello, Edoardo and Kovachki, Nikola B and Levine, Matthew E and Stuart, Andrew M},
  journal={arXiv preprint arXiv:2406.06486},
  year={2024}
}

@article{torlai2018neural,
  title={Neural-network quantum state tomography},
  author={Torlai, Giacomo and Mazzola, Guglielmo and Carrasquilla, Juan and Troyer, Matthias and Melko, Roger and Carleo, Giuseppe},
  journal={Nature physics},
  volume={14},
  number={5},
  pages={447--450},
  year={2018},
  publisher={Nature Publishing Group UK London}
}

@article{jia2019quantum,
  title={Quantum neural network states: A brief review of methods and applications},
  author={Jia, Zhih-Ahn and Yi, Biao and Zhai, Rui and Wu, Yu-Chun and Guo, Guang-Can and Guo, Guo-Ping},
  journal={Advanced Quantum Technologies},
  volume={2},
  number={7-8},
  pages={1800077},
  year={2019},
  publisher={Wiley Online Library}
}

@article{gros1993cluster,
  title={Cluster expansion for the self-energy: A simple many-body method for interpreting the photoemission spectra of correlated Fermi systems},
  author={Gros, Claudius and Valenti, Roser},
  journal={Physical Review B},
  volume={48},
  number={1},
  pages={418},
  year={1993},
  publisher={APS}
}

@article{pfau2020ab,
  title={Ab initio solution of the many-electron Schr{\"o}dinger equation with deep neural networks},
  author={Pfau, David and Spencer, James S and Matthews, Alexander GDG and Foulkes, W Matthew C},
  journal={Physical review research},
  volume={2},
  number={3},
  pages={033429},
  year={2020},
  publisher={APS}
}

@article{hou2024unsupervised,
  title={Unsupervised representation learning of Kohn--Sham states and consequences for downstream predictions of many-body effects},
  author={Hou, Bowen and Wu, Jinyuan and Qiu, Diana Y},
  journal={Nature Communications},
  volume={15},
  number={1},
  pages={9481},
  year={2024},
  publisher={Nature Publishing Group UK London}
}

@article{boulle2023mathematical,
  title={A Mathematical Guide to Operator Learning},
  author={Boull{\'e}, Nicolas and Townsend, Alex},
  journal={arXiv preprint arXiv:2312.14688},
  year={2023}
}

@article{zhu2025predicting,
  title={Predicting nonequilibrium Green’s function dynamics and photoemission spectra via nonlinear integral operator learning},
  author={Zhu, Yuanran and Yin, Jia and Reeves, Cian C and Yang, Chao and Vl{\v{c}}ek, Vojt{\v{e}}ch},
  journal={Machine Learning: Science and Technology},
  volume={6},
  number={1},
  pages={015027},
  year={2025},
  publisher={IOP Publishing}
}

@article{liang2024effective,
  title={Effective many-body interactions in reduced-dimensionality spaces through neural network models},
  author={Liang, Senwei and Kowalski, Karol and Yang, Chao and Bauman, Nicholas P},
  journal={Physical Review Research},
  volume={6},
  number={4},
  pages={043287},
  year={2024},
  publisher={APS}
}

@article{liang2025exploring,
  title={Exploring the Nexus of Many-Body Theories through Neural Network Techniques: the Tangent Model},
  author={Liang, Senwei and Kowalski, Karol and Yang, Chao and Bauman, Nicholas P},
  journal={arXiv preprint arXiv:2501.15792},
  year={2025}
}

@article{kakizawa2024physics,
  title={Physics-informed neural network model for quantum impurity problems based on Lehmann representation},
  author={Kakizawa, Fumiya and Terasaki, Satoshi and Shinaoka, Hiroshi},
  journal={arXiv preprint arXiv:2411.18835},
  year={2024}
}

@article{agapov2024predicting,
  title={Predicting interacting Green's functions with neural networks},
  author={Agapov, Egor and Bertomeu, Oriol and Carballo, Andr{\'e}s and Mendl, Christian B and Sander, Aaron},
  journal={arXiv preprint arXiv:2411.13644},
  year={2024}
}

@article{venturella2024unified,
  title={Unified Deep Learning Framework for Many-Body Quantum Chemistry via Green's Functions},
  author={Venturella, Christian and Li, Jiachen and Hillenbrand, Christopher and Peralta, Ximena Leyva and Liu, Jessica and Zhu, Tianyu},
  journal={arXiv preprint arXiv:2407.20384},
  year={2024}
}

@article{blankenbecler1981monte,
  title={Monte Carlo calculations of coupled boson-fermion systems. I},
  author={Blankenbecler, Richard and Scalapino, DJ and Sugar, RL},
  journal={Physical Review D},
  volume={24},
  number={8},
  pages={2278},
  year={1981},
  publisher={APS}
}

@article{antipov2017currents,
  title={Currents and Green's functions of impurities out of equilibrium: Results from inchworm quantum Monte Carlo},
  author={Antipov, Andrey E and Dong, Qiaoyuan and Kleinhenz, Joseph and Cohen, Guy and Gull, Emanuel},
  journal={Physical Review B},
  volume={95},
  number={8},
  pages={085144},
  year={2017},
  publisher={APS}
}

@article{fei2021analytical,
  title={Analytical continuation of matrix-valued functions: Carath{\'e}odory formalism},
  author={Fei, Jiani and Yeh, Chia-Nan and Zgid, Dominika and Gull, Emanuel},
  journal={Physical Review B},
  volume={104},
  number={16},
  pages={165111},
  year={2021},
  publisher={APS}
}

@article{maier2005quantum,
  title={Quantum cluster theories},
  author={Maier, Thomas and Jarrell, Mark and Pruschke, Thomas and Hettler, Matthias H},
  journal={Reviews of Modern Physics},
  volume={77},
  number={3},
  pages={1027--1080},
  year={2005},
  publisher={APS}
}

@article{vaswani2017attention,
  title={Attention is all you need},
  author={Vaswani, Ashish and Shazeer, Noam and Parmar, Niki and Uszkoreit, Jakob and Jones, Llion and Gomez, Aidan N and Kaiser, {\L}ukasz and Polosukhin, Illia},
  journal={Advances in neural information processing systems},
  volume={30},
  year={2017}
}

@inproceedings{hao2023gnot,
  title={Gnot: A general neural operator transformer for operator learning},
  author={Hao, Zhongkai and Wang, Zhengyi and Su, Hang and Ying, Chengyang and Dong, Yinpeng and Liu, Songming and Cheng, Ze and Song, Jian and Zhu, Jun},
  booktitle={International Conference on Machine Learning},
  pages={12556--12569},
  year={2023},
  organization={PMLR}
}

\end{document}